%% file: FPoD.tex
\documentclass[numberedappendix, iop, apj]{emulateapj}
\usepackage{natbib}
\usepackage{url}
\usepackage{mathtools}

\DeclareMathAlphabet{\mathpzc}{OT1}{pag}{m}{rm}

\def\lya{Ly-$\alpha$}

\def\nh{$N_{\text{H\,\tiny{I}}}$}

\def\delv{$\Delta$$\rm{v_{90}}$}

\def\elc{$\ell_{\rm{c}}$}
\def\w1526{$W_{1526}$}
\def\delvres{$\Delta$v$_{\rm{res}}$}

\def\kms{km~s$^{-1}$}
\def\delvmax{$\Delta$$\rm{v_{90,max}}$}
\def\hi{\ion{H}{1}}

\newcommand{\noprint}[1]{}

\slugcomment{submitted to ApJ 2012 December 5; accepted 2013 April 4; published 2013 May 3}

\shorttitle{The Fundamental Plane of Damped {\lya} Systems}
\shortauthors{Neeleman et al.}

\begin{document}

\title{The Fundamental Plane of Damped {\lya} Systems}


\author{Marcel Neeleman, Arthur M. Wolfe}
\affil{Department of Physics and Center for Astrophysics and Space Sciences, UCSD, La Jolla, CA 92093, USA}
\email{mneeleman@physics.ucsd.edu}
\author{J. Xavier Prochaska}
\affil{Department of Astronomy \& Astrophysics, UCO/Lick Observatory, 1156 High Street, \\
University of California, Santa Cruz, CA 95064, USA}
\and
\author{Marc Rafelski}
\affil{Infrared Processing and Analysis Center, Caltech, Pasadena, CA, 91125, USA}

\begin{abstract} 
Using a sample of 100 {\hi} - selected damped Ly-$\alpha$ (DLA) systems, observed with the High Resolution Echelle Spectrometer on the Keck I telescope, we present evidence that the scatter in the well-studied correlation between the redshift and metallicity of a DLA is largely due to the existence of a mass-metallicity relationship at each redshift. To describe the fundamental relations that exist between redshift, metallicity and mass, we use a fundamental plane description, which is described by the following equation: $[\rm{M/H}]=(-1.9{\pm}0.5)+(0.74{\pm}0.21){\cdot}{\log}{\Delta}v_{90}-(0.32{\pm}0.06){\cdot}z$. Here, we assert that the velocity width, {\delv}, which is defined as the velocity interval containing 90\% of the integrated optical depth, traces the mass of the underlying dark matter halo. This description provides two significant improvements over the individual descriptions of the mass-metallicity correlation and metallicity-redshift correlation. Firstly, the fundamental equation reduces the scatter around both relationships by about 20\%, providing a more stringent constraint on numerical simulations modeling DLAs. Secondly, it confirms that the dark matter halos that host DLAs satisfy a mass-metallicity relationship \emph{at each redshift} between redshifts 2 through 5.\\
\end{abstract}


\section{Introduction}
\label{sec:intro}
The study of high-redshift quasars has resulted in the detection of over 6000 damped Lyman $\alpha$ absorption systems (DLAs), which are absorption systems with an {\hi} column density greater than $2 \times 10^{20}$ cm$^{-2}$ \citep{Prochaska2005, Prochaska2009, Noterdaeme2009, Noterdaeme2012}. DLAs are thought to be the progenitors of modern day galaxies \citep{Wolfe1995}, and therefore provide an important observational constraint on the formation and evolution of galaxies seen today. Most DLAs are found using low resolution instruments such as the Sloan Digital Sky Survey \citep[SDSS;][]{Abazajian2009}. Although these instruments allow for accurately determining the column density of the neutral hydrogen, follow-up measurements of DLAs using high resolution instruments are needed to accurately measure the resonance line transitions of metals that are present in the DLAs. These measurements yield the column densities of the DLA's gas-phase constituents \citep[for a review see:][]{Wolfe2005}, and allow for the determination of the metallicity of the gas \citep[e.g.][]{Prochaska2003b}, the cooling rate of the neutral gas due to [\ion{C}{2}] 158 $\mu$m emission \citep[{\elc};][]{Wolfe2003}, and the dust-to-gas ratio \citep{Pettini1994}. Besides these quantities, the velocity profiles of the metal lines also gives us an insight into the kinematics of the gas \citep[e.g.][]{Prochaska1997, Ledoux1998}. Together with the redshift of the DLA, these parameters individually provide us insight into the composition and nature of DLAs through the distribution that each of these parameters assumes. In turn, these data offer insight into the properties of gas-rich and presumably starforming galaxies at high redshift. Furthermore, we may obtain additional information by considering how these parameters correlate to each other.

There are two such correlations for DLAs that have been extensively examined. The first of these correlations is the evolution of metallicity with redshift; as redshift decreases, metallicity increases \citep{Prochaska2003a, Kulkarni2005, Rafelski2012}. The second is the correlation between the kinematics and the metallicity of the DLA \citep{Wolfe1998, Ledoux2006, Prochaska2008}. This second relationship is thought to be caused by the underlying mass-metallicity relationship seen in galaxies \citep{Tremonti2004, Savaglio2005, Erb2006, Maiolino2008}, where the kinematics of the DLA is assumed to be related to the virial speed, hence the mass of the dark matter halo that hosts the DLA. \citet{Fynbo2010, Fynbo2011} used this correlation to image DLAs in emission. By selecting higher metallicity systems, they improved their rate of detection, which they assert is due to the larger mass of the DLA.

One of the main features of both correlations is the large intrinsic scatter, which is significantly larger than the observational uncertainty in the measurements. One possible cause for the large intrinsic scatter around either correlation is that the two correlations are dependent on each other, since they both correlate metallicity to a second parameter. To quantify the potential dependency of the two correlations, we can see if a linear combination of the two correlations is able to reduce the intrinsic scatter. Such a linear combination is called a `fundamental' plane, because it traces out a plane in the three dimensional parameter space it describes. 

The occurrence of fundamental planes in astronomy is relatively common. Indeed, in recent years there have been several publications on the existence of fundamental planes in a variety of astronomical areas, from black hole activity \citep{Merloni2003}, to that of radio magnetars \citep{Rea2012}. To illustrate the advantage of using fundamental planes, we will consider one of the first fundamental planes, the fundamental plane of elliptical galaxies. Elliptical galaxies show a great diversity in their properties such as surface brightness and velocity dispersion. However, most of these properties are correlated to the luminosity of the elliptical galaxy \citep{Kormendy1982}. Indeed, both velocity dispersion and surface brightness show a correlation with luminosity. However, the scatter around both correlations is larger than the observational uncertainty. In an effort to reduce this scatter, \citet{Djorgovski1987} noticed that when the luminosity, velocity dispersion and surface brightness are plotted on a three-dimensional plot, the points line up on a plane with a scatter around the plane that is within the observational uncertainty. This planar description therefore showed that the majority of the scatter in either of the correlations was caused by the other correlation. As such, the plane description provides a more stringent constraint on theoretical models than each of the individual correlations. In addition, the coefficients that describe this plane provide us insight into the galaxy population as a whole; deviations from the expected values could indicate variations in the initial mass function, dark matter fraction and stellar population effects \citep[see e.g.][]{Graves2010}. 

In this paper, we explore the possibility of combining the redshift-metallicity and mass-metallicity correlation into a `fundamental' plane for DLAs. We find that indeed such a plane exists, and applying this plane decreases the intrinsic scatter seen in both correlations significantly. We discuss new insights provided by this fundamental plane and we explore the physical origin of such a plane.

We conclude the paper by looking for the existence of this fundamental plane in current numerical simulations. Since most numerical simulations do not accurately trace all parameters involved; direct evidence of the fundamental plane is difficult to find. Instead we focus on projections of the fundamental plane; i.e. we primarily compare two parameter correlations and distributions of a single parameter. One distribution in particular that we examine is the velocity width distribution. This distribution has two notable features, a very large median and a tail toward large velocity widths. Previous modeling showed that only two models were able to recreate these two features of the distribution. The first model posed that DLAs are \emph{thick} rapidly rotating disks \citep{Prochaska1997}, whereas the second model posed that DLAs are disk progenitors in a standard cold dark matter cosmology \citep{Haehnelt1998}. The latter, however, erroneously excluded systems with low velocity widths \citep{Prochaska2010}. Moreover, later numerical simulations based on $\Lambda$CDM models that included a more realistic transport of ionizing radiation \citep{Razoumov2006, Pontzen2008} were unable to recreate the distribution of observed velocity widths. This inability of simulations based on $\Lambda$CDM models to produce the observed velocity width distribution is attributed to their difficulty in producing enough DLAs that reside in dark matter halos with large masses \citep{Pontzen2008}. We explore this hypothesis with the current data set.

This paper is organized as follows. In Section \ref{sec:sample} we describe our observational strategy and the reduction process we used. We explain the method used to measure the parameters described for each DLA, consider any biases in the sample, and explore their effects on the distribution of the parameters. Since a fundamental plane is dependent on the correlations between the parameters involved, it is crucial to explore these correlations. This is done in Section \ref{sec:distandcor}. In Section \ref{sec:fund} we introduce the fundamental plane. Finally, we will discuss these results and compare them to numerical models in Section \ref{sec:disc}. The sample discussed in this paper is the largest sample of DLAs for which the metal lines have been observed with a single high resolution (R $>$ 40000) instrument \citep[Keck/HIRES;][]{Vogt1994}, and to our knowledge, this is the first paper that considers multi-parameter correlations, such as fundamental planes, for DLAs.

\section{Sample Selection, Parameter Definition and Sample Biases}
\label{sec:sample}

In this section we discuss the reduction process used for the sample. We explain our sample selection and any biases that could influence our results.

\input{./table1.tex}

\subsection{Observations and Reduction}
All of the spectra in this sample were obtained with the High Resolution Echelle Spectrometer \citep[HIRES:][]{Vogt1994} on the Keck I 10m telescope over the course of almost two decades of observing. Table \ref{tab:obs} shows a detailed journal of observations. All observations prior to 2004 were carried out with the Tektronix 2048x2048 CCD; the remaining observations were made with a 3-chip mosaic of MIT-LL 2048x4096 CCDs. The FWHM resolution of each object, {\delvres}, varies depending on the specified slit size and the atmospheric conditions, but most of the data were taken with a 0.86$\arcsec$ or 1.15$\arcsec$ slit, which results in a maximum {\delvres} corresponding to 6 {\kms}, and 8 {\kms} respectively. Wavelength coverage was dependent on the redshift of the DLA. Finally, we computed the average signal to noise ratio (S/N) per 1.4 {\kms} pixel for each observation, by taking the central 200{\AA} of each observation and finding the median S/N for this range. If the range included too many absorption features (e.g. atmospheric waterlines, etc.) we shifted the range by 200{\AA} and computed the median S/N for this range. The raw data were reduced using the $\tt{HIRedux}$ routine, then extracted, coadded and continuum fit with $\tt {x\_continuum}$. These routines are all part of the publicly available $\tt XIDL$ reduction package developed by J.X. Prochaska \citep{Prochaska2003b}. The reasons for choosing a single high resolution (i.e. R $>$ 40000) instrument are discussed in \S\ref{sec:bias}.

\subsection{Methodology}
\label{param_def}
After the reduction process, we searched the reduced spectra for metal lines at the redshift of the DLA obtained from the Lyman alpha line in lower resolution data. The redshift of the DLA was then adjusted to coincide with the peak absorption feature in the metal lines, since metal lines are much narrower than the large width of the damped profile of the Ly$\alpha$ line. This provides a more accurate measurement of the DLA redshift. We also calculated the column densities of metals using the apparent optical depth method \citep[AODM;][]{Savage1991}, which yields accurate column densities even if some of the lines are slightly saturated. From the column densities of the metal lines and the column density of hydrogen, we can calculate the metallicity ([M/H]) of the DLA, which is defined by:
\begin{equation}
[\rm{M/H}]=\log_{10}(N_{\rm{M}}/N_H)-\log_{10}(N_{\rm{M}}/N_H)_{\odot}
\end{equation}
The wavelengths and oscillator strengths used in this paper are from \citet{Morton2003}, whereas the solar abundances are from \citet{Asplund2009}. 

Besides the metallicity we can also determine the cooling rate of neutral gas from these column densities. \citet{Pottasch1979} showed that, {\elc}, the cooling rate per H atom due to [\ion{C}{2}] 158 $\mu$m emission, which is the dominant coolant in the neutral ISM \citep{Wright1991}, is given by:
\begin{equation}
\ell_{\rm{c}}=\frac{N(\text{\ion{C}{2}}^*)}{N_{\text{H\,\tiny{I}}}}A_{ul}h\nu_{ul},
\end{equation}
where $A_{ul}$ and $h\nu_{ul}$ are the Einstein coefficient and energy of the transition from the excited to the ground state of C$^+$ \citep[see e.g.][]{Wolfe2003}. We are able to measure the column density of \ion{C}{2}$^*$, because the \ion{C}{2}$^*$$\lambda$1335.7 fine structure line arises from the same state as the [\ion{C}{2}] 158 $\mu$m line, and falls within the spectral regime covered by optical telescopes for the redshifts examined in this paper. 

Note that an accurate {\hi} column density, {\nh}, is required to obtain accurate measurements of both metallicity and {\elc}. We are generally unable to obtain the {\hi} column density from the HIRES spectra, because accurate measurement of this quantity requires the spectrum to be fluxed and HIRES spectra are difficult to flux \citep{Suzuki2003}. We therefore relied mainly on spectra obtained from the Sloan Digital Sky Survey \citep[SDSS;][]{Abazajian2009} to measure {\nh}. However, for those 41 DLAs that were observed with the Echelette and Imaging Spectrometer \citep[ESI;][]{Sheinis2002} on the Keck II 10m telescope, we used the ESI spectra, because the improved resolution provides more reliable and precise {\hi} column densities. The column density of {\hi} is determined from these spectra by simultaneously fitting a Voigt profile to the absorption profile of the Ly$\alpha$ line of the DLA, and fitting the continuum of the fluxed spectrum of the background quasar \citep[see e.g.][]{Prochaska2003a, Rafelski2012}. 

The final two quantities discussed here are the two kinematic parameters, {\w1526} and {\delv}. The reason for using two different kinematic parameters is that they describe different kinematical properties of the DLA. The {\delv} parameter describes the velocity width of the main neutral absorption complex by specifically ignoring weak outlying velocity components. On the other hand, the {\w1526} parameter is dominated by these outliers, since this absorption line is saturated in most cases \citep{Prochaska2008}. Hence the {\w1526} parameter describes the kinematics of the halo gas and/or weak satellite components.

The rest equivalent width of the \ion{Si}{2}$\lambda$1526 line, {\w1526}, is defined as $W_{1526}=W_{obs}/(1+z)$, where $W_{obs}$ is the observed equivalent width of the \ion{Si}{2}$\lambda$1526 transition. We choose the \ion{Si}{2}$\lambda$1526 line because the line is measured in the majority of spectra and its high oscillator strength causes the line to be saturated in most systems. It is important to stress that the {\w1526} statistic is almost independent of the column density of Si$^+$,$N_{\rm{Si^+}}$, \emph{if} the absorption line is saturated. Saturated lines are on the flat part of the curve of growth where the column density and equivalent width scale as, $W_{1526}{\propto}\sqrt{\ln{N_{\rm{Si^+}}}}$; any reasonable change in $N_{\rm{Si^+}}$ would only marginally change $W_{1526}$. There is therefore no \emph{a priori} strong correlation expected between the equivalent width and any parameter which is derived from the column density of any metal lines, such as metallicity, if the equivalent width is obtained from a saturated line. Consequently, we were mindful to only select those DLAs with saturated \ion{Si}{2}$\lambda$1526 transition lines for the comparison of the {\w1526} parameter with any other parameter. However, the unsaturated lines were included in the distribution for {\w1526}; otherwise we would bias this distribution towards higher equivalent widths, since low equivalent width systems are more likely to be unsaturated due to the mass-metallicity relationship.

The velocity width, {\delv}, of an absorption system is defined to be the width of the absorption profile in velocity space. To measure the velocity width of a DLA, we employ the same strategy as in \citet{Prochaska1997}, with a few exceptions. We first select an unsaturated low-ion transition line. We require the line to be unsaturated, because a saturated line could overestimate the size of the velocity interval by including weak outlying velocity features that contain little of the total neutral gas of the DLA. As such we require that the normalized flux, defined as $F_{\rm{norm}}=I(\rm{v})/I_c$, where $I_{c}$ is the continuum intensity incident on the gas and $I(\rm{v})$ the transmitted intensity, is greater than 0.1 over the entire absorption profile. A low-ion transition line is chosen because low-ions, such as \ion{Fe}{2}$\lambda$1608 and \ion{Si}{2}$\lambda$1808, are likely to trace the neutral gas, which creates the damped Lyman $\alpha$ profile \citep{Wolfe1995, Prochaska1997}. By contrast, higher ionization transition lines such as \ion{C}{4}${\lambda}{\lambda}$1548, 1550  and \ion{Si}{4}${\lambda}{\lambda}$1393, 1402 exhibit different velocity structures \citep{Wolfe2000, Fox2007}. In addition, we require that the line is unblended from any other absorption feature.

\begin{figure}
\epsscale{1.09}
\plotone{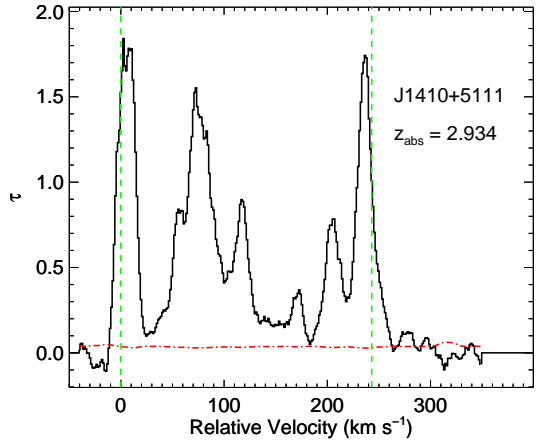}
\caption{Optical depth profile as a function of relative velocity. The sample profile shown is of the DLA at redshift $z_{\rm{abs}}=2.9340$ towards QSO J1410+5111. The profile has been smoothed to 8~{\kms}, which is the largest {\delvres} used in this sample. The dashed (green) lines mark the {\delv} interval, which is 243~{\kms}, and the dashed-dotted (red) line is the smoothed 1-$\sigma$ uncertainty on the data. For the whole sample, the {\delv} values range between 17~{\kms} and 484~{\kms}.}
\label{fig:od}
\end{figure}

After selecting the absorption profiles, we obtain an apparent optical depth profile;
\begin{equation}
\tau(\nu)=\ln [{F_{\rm{norm}}}^{-1}].
\label{eq:lc}
\end{equation}
The resultant profile is smoothed to 8 {\kms}, the largest {\delvres} of our sample, to prevent differences in resolution from affecting the {\delv} values. After smoothing, we select only those profiles for which the peak optical depth is detected at the 12$\sigma$ detection limit, so that components one-fourth this peak optical depth are detected with a 3$\sigma$ detection limit. This is important because these components could contain a significant fraction of the neutral gas content. This criterion is less restrictive than the criterion used by \citet{Prochaska1997}, because we believe that a 3$\sigma$ detection limit is enough to discern small absorption features, since in almost all cases we have stronger absorption lines that clearly show the presence of these small absorption features above the 5$\sigma$ detection limit, and we are only after the velocity width of the line and not any of the other parameters described in \citet{Prochaska1997}. Finally, we find the width of the profile by stepping inward from both sides of the profile pixel-by-pixel, until we reach 5 percent of the total integrated optical depth on each side. This width is the measured {\delv} value. This last step prevents weak outlying absorption features from skewing the {\delv} statistic. The complete set of smoothed optical depths as a function of relative velocity for all DLAs is shown in Figure \ref{fig:od}. For display purposes, we shift the data such that the left edge of the profile lines up with 0 {\kms}. The smoothed 1-$\sigma$ error array is shown as a (red) dashed dotted line. The separation between the (green) dashed lines marks the velocity width. We assume an error of 10 {\kms} on these measurements, similar to that used by \citet{Prochaska2008}.

\input{./table2.tex}

\subsection{Systematic Errors and Biases due to Sample Selection Criteria}
\label{sec:bias}
There are two main concerns in selecting an appropriate sample for exploring correlations: systematic errors and biases caused by sample selection criteria. One source of systematic error could be the use of two different resolution instruments, which would in particular affect the velocity width measurement since it is measured by moving in pixel-by-pixel in the spectra. Lower resolution spectra, like the spectra taken with ESI which has a resolution of 44 {\kms}, would overestimate the velocity width. This motivated \citet{Prochaska2008} to reduce their ESI {\delv} measurements by 20 {\kms}. To circumvent the systematic errors caused by multiple instruments, we decided to restrict ourselves to velocity width measurements and metal column densities obtained from HIRES alone.

The other main concern is the effect of sample selection biases on the data. To assess this effect, we recorded the reason why each DLA was observed by our group with HIRES. The reason generally falls in one of the following four categories: (1) {\hi}-selected, the DLA was selected based on just the column density of {\hi}; (2) high redshift, the DLA was selected because of its high redshift; (3) metal content, the DLA exhibited very strong metal lines in its spectrum obtained from SDSS; and finally, (4) serendipitous, the DLA happened to fall in the same line-of-sight of the primary target. In addition, the majority of targets were observed because the flux of the background quasar exceeded a certain threshold; although the second and third reasons often drove our group to observe fainter targets. Each of these selection criteria needs to be examined for any bias that the corresponding subsample might contain. First, if the distance between the DLA and quasar is sufficient, then the DLA properties are unrelated to the properties of the background quasar. This indicates that selecting DLAs by the flux of the background quasar should not bias the DLA sample. Similarly for the serendipitous subsample, if the separation between the primary target and the serendipitous DLA is large enough, the serendipitous DLA should be unrelated to the primary target. We take a velocity separation of 3000 {\kms} to be sufficient for a DLA to be unaffected by another DLA or quasar \citep[see e.g.][]{Ellison2010}, and we have made sure that no DLA in our sample violates this criterion. Secondly, the DLAs that were observed because of their metal column density \citep[metal-strong DLAs, MSDLAs;][]{Herbert-Fort2006} are not included in our sample. This subsample is biased in metallicity and {\w1526}, because metal-selected DLAs have on average higher metallicities and correspondingly higher {\w1526} values \citep{Kaplan2010}.

This leaves two sample selection criteria; the redshift and the {\hi} column density of the DLA; both of these sample selection criteria cause a bias in the sample. The bias in the redshift is twofold. First, bright quasars are more abundant in the redshift range 2-3, and therefore any magnitude-limited selection of quasars will produce a majority of DLAs within this range \citep{Prochaska2005, Prochaska2009}. Second, for a subset of the data we selected the quasars based on their high redshift, which biases that sample towards DLAs with high redshift. The {\hi} column density selection criteria causes our sample to contain proportionally too few DLAs below a column density of $10^{20.5} \rm{cm}^{-2}$ compared to the sample of DLAs from SDSS. \citet{Prochaska2007} ascribe this to our preference in selecting higher column density systems to ensure the absorber satisfies the DLA  {\hi} column density criterion. 

To gauge the extent of the bias on the distribution of the remaining four parameters; we want to compare our biased sample to one that is `free' of the bias caused by the redshift and {\hi} column density selection criteria. To accomplish this, we randomly divide our sample of 100 DLAs in half. One half is left untouched, whereas we randomly pick 50 DLAs (with repeats) from the second half such that the redshift and {\hi} column density of the second half reproduces the {\hi} column density frequency distribution observed in DLAs, \citep[$f_{\text{H\,\tiny{I}}}(N,X)$;][]{Prochaska2005, Prochaska2009, Noterdaeme2009, Noterdaeme2012}. This is indicative of a sample that is `free' of biases in redshift and {\hi} column density. To check for a bias in metallicity, {\delv}, {\w1526}, and {\elc} of the untouched half, we look at the distribution that these parameters assume for both halves of the data set. We compare the distributions of each half, using a two-sided Kolmogorov-Smirnov test (KS) test and a Mann-Whitney U-test. The KS test provides a probability that both populations are drawn from the same parent population, whereas the U-test provides a probability that the two median values are significantly different. This procedure of randomly dividing the sample, reselecting one half of the sample and comparing the distribution is repeated 1000 times, to get a median value for the results of both tests.

The resultant median values of the KS test for all parameters are greater than 0.05 indicating that we cannot rule out the null hypothesis, which is that the two subsamples were drawn from the same parent population, at a 95 \% confidence level (c.l.). Similarly, the U-test yields mean values greater than 0.05 for all parameters indicating that we also cannot reject this null hypothesis, which is that the medians of the two subsamples are the same. Hence, we see no evidence that the redshift and {\hi} column density selection criteria have significantly affected the distribution of the remaining parameters. It is important to note that this does \emph{not} mean that the redshift and {\hi} column density show no correlation with the remaining parameters. It means that the effect of the two sample selection criteria does not significantly affect the distributions of the remaining four parameters. Together with the serendipitous sample, the {\hi}-selected and redshift selected subsample comprises the complete sample used in this paper. In total, it consists out of 100 DLAs for which we have accurate measurements of metallicity, redshift, {\hi} column density, and {\delv}; the full sample is shown in Table \ref{tab:dat}.

\section{Distributions and Correlations}
\label{sec:distandcor}

In this section, we compare the distribution of metallicity, {\elc}, and {\w1526} parameters to previous studies specifically aimed at exploring these parameters, to show that our sample of 100 DLAs is a representative sample of each of the parameters. We explore our sample of 100 {\delv} measurements of DLAs, which to our knowledge is the most accurate distribution to date of this quantity. We also look at any potential correlations between the parameters, except for those that have been explored in detail before, such as the kinematics-metallicity correlation \citep{Wolfe1998, Ledoux2006, Prochaska2008} and the metallicity-redshift correlation \citep{Prochaska2003a, Kulkarni2005, Rafelski2012}.

\begin{figure*}
\epsscale{1.07}
\plottwo{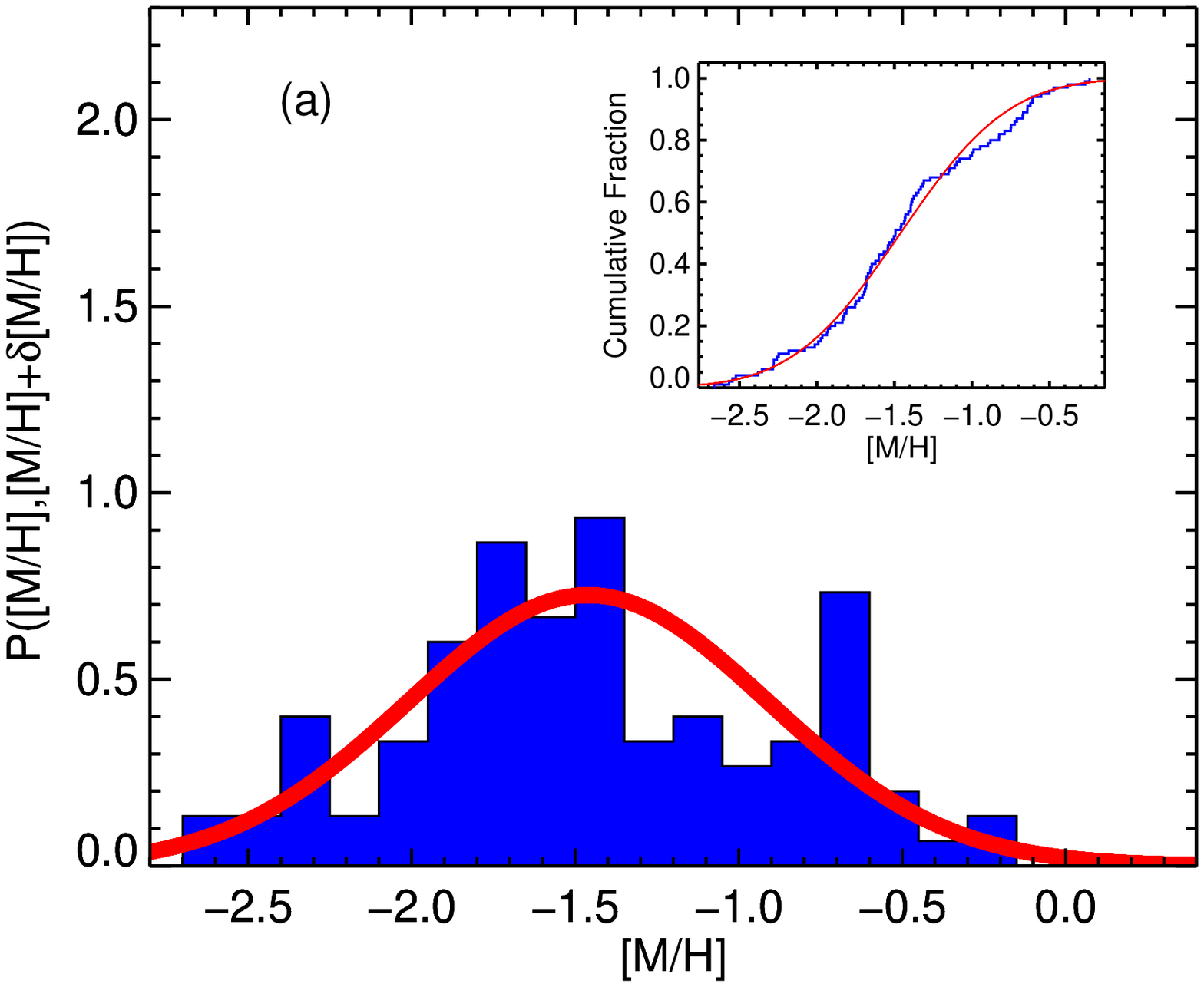}{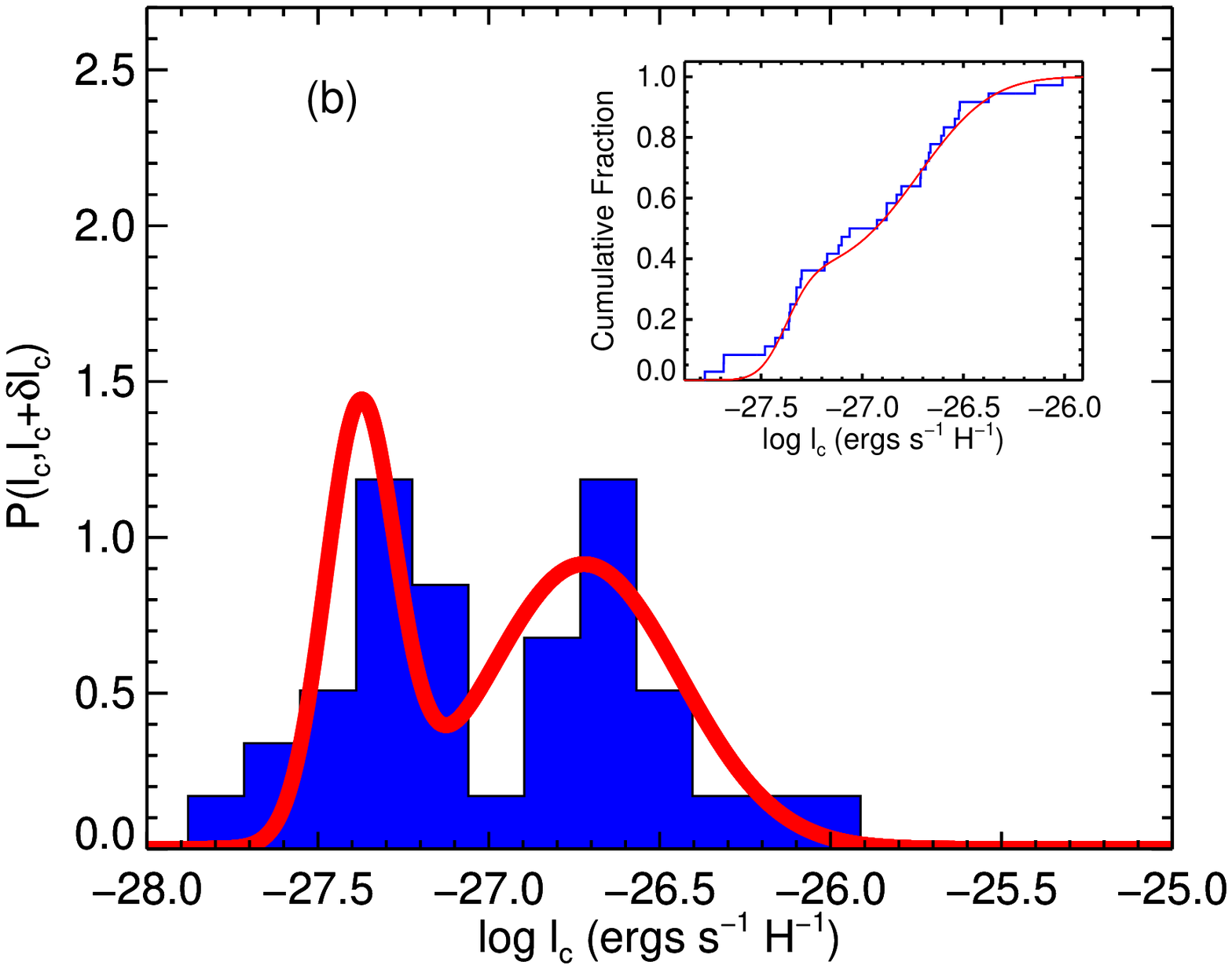}
\plottwo{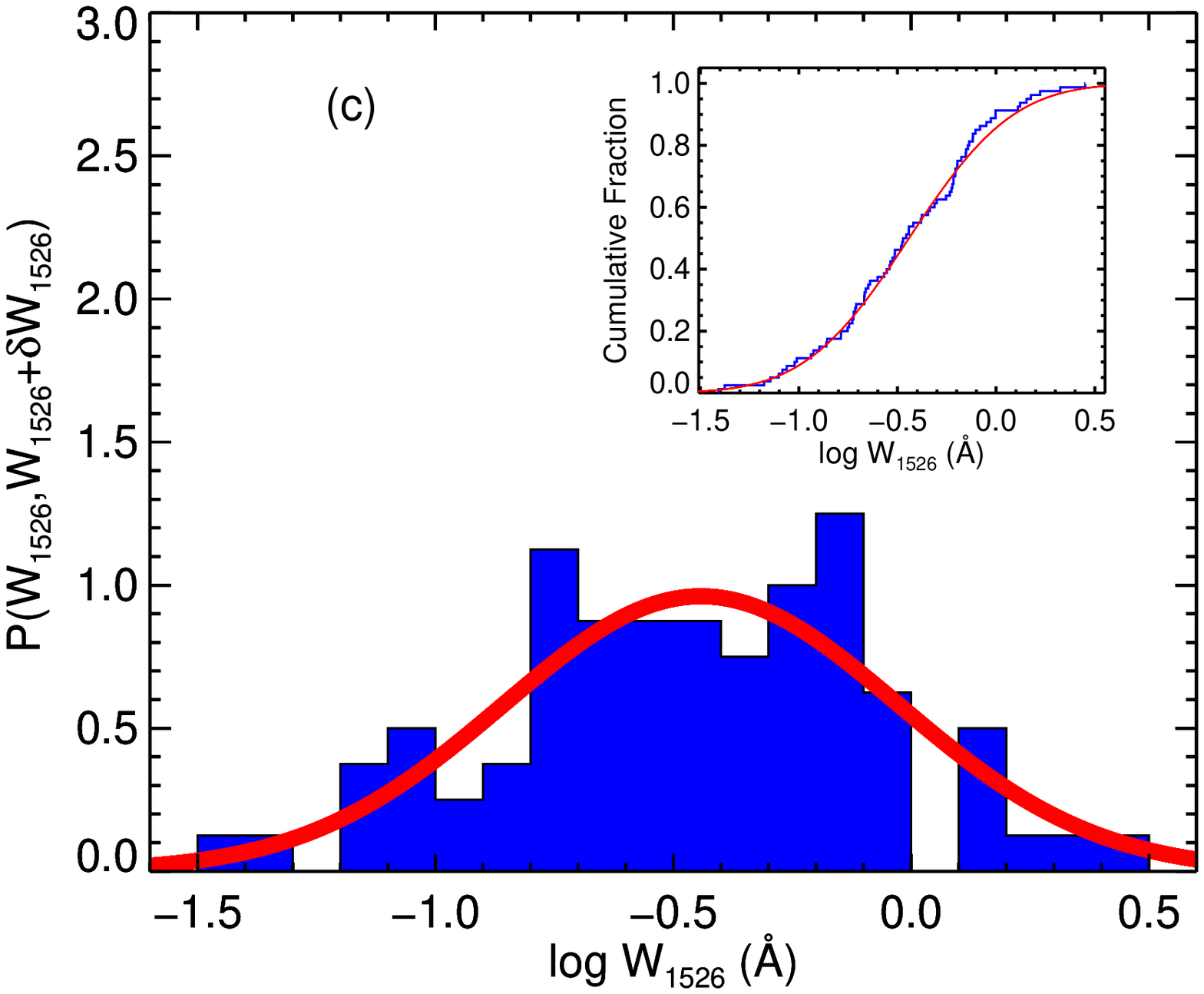}{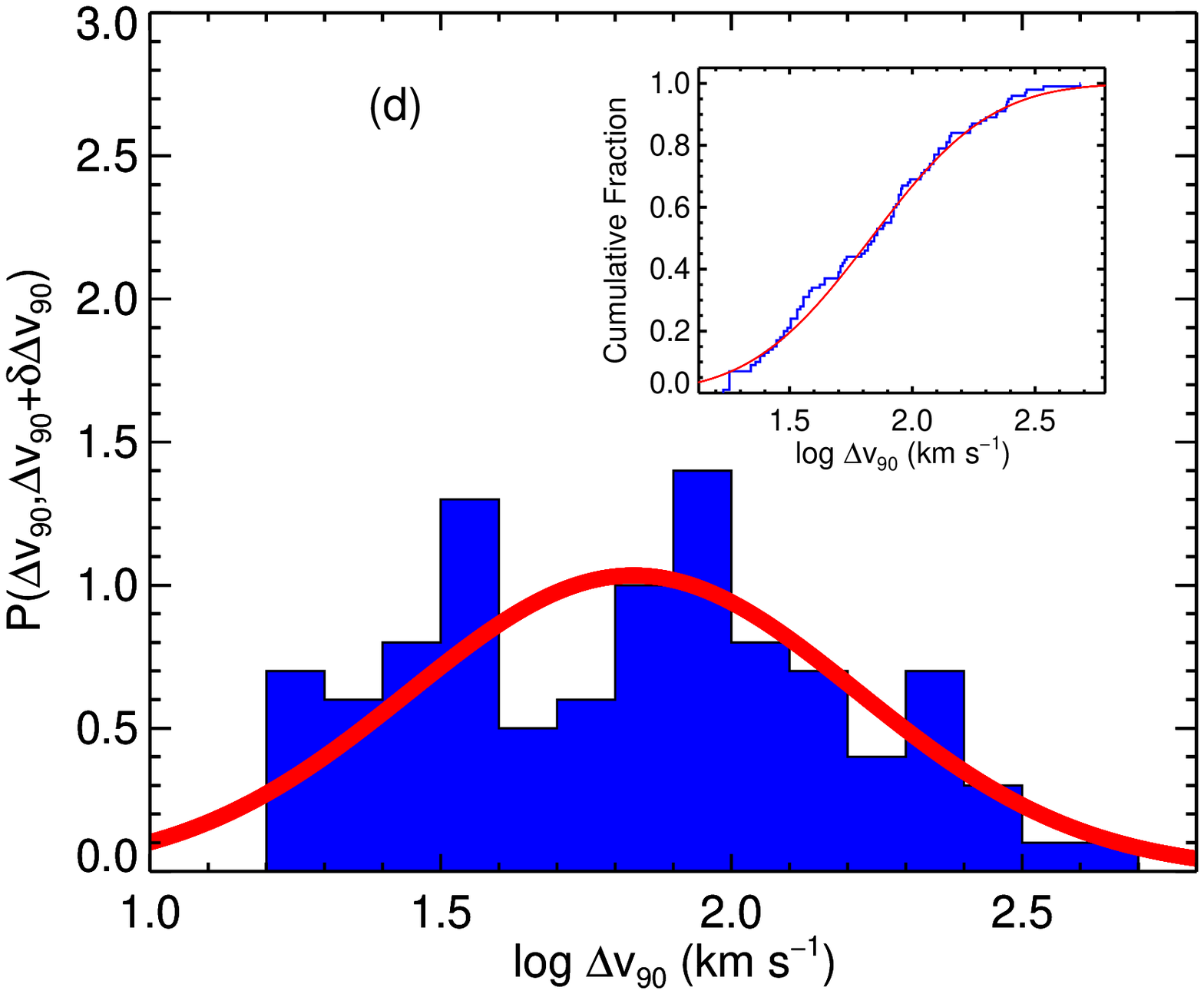}
\caption{Distributions of metallicity, 158 $\mu$m cooling rate, equivalent width and velocity width of the sample of DLAs. The solid (red) line is a chi-squared best fit line to the cumulative distributions (shown as insets in the figure); the equations of these best fit parameters are given in Table \ref{tab:fit}. The distributions of metallicity, {\delv} and {\w1526} can all be described by a single Gaussian distribution. The {\elc} parameter on the other hand, is clearly bimodal and is best described by a bimodal distribution. The distributions of metallicity, {\elc} and {\w1526} are in good agreement with previous studies, indicating that the sample is a representative subsample of each of these parameters. The {\delv} distribution shows a large number of DLAs with {\delv} $>$ 100 {\kms}, which raises the median {\delv} value to 70$_{-13}^{+16}$ {\kms}.}
\label{fig:dist}
\end{figure*}

\subsection{Distributions}
\label{sec:dist}
The distributions of metallicity, {\elc}, and {\w1526} of the sample are shown in Figure \ref{fig:dist}(a-c). These distributions have been discussed in detail in previous papers and here we give a short summary of the characteristics of each distribution. First, we scale each distribution such that its integrated area is normalized to unity. Consequently, we can interpret the y-axis of parameter, $p$, as the probability of finding a DLA within the parameter range $(p, p+dp)$. We then fit analytic functions to each of the distributions described above; these are shown as the solid (red) line in Figure \ref{fig:dist}. To prevent binning from affecting the fit, we do not fit the actual distributions, but instead fit the cumulative distribution function (shown as insets in Figure \ref{fig:dist}). The procedure used for the fitting process is as follows. We first use a chi-squared fitting technique to fit the cumulative distribution function of a single Gaussian distribution function to the cumulative distribution of the data; this fit is then compared using a one-sided KS-test. If the fit is unacceptable, we try for the fitting function a sum of two Gaussian distributions. The resultant equations to the fit are shown in Table \ref{tab:fit}. As expected, the {\elc} statistic is best described by a bimodal fitting function \citep{Wolfe2008}, whereas the other parameters can be described by a single Gaussian distribution.

\input{./table3.tex}

Figure \ref{fig:dist}a shows the metallicity distribution, which was recently published by \citet{Rafelski2012}. They show that the distribution is reasonably well described by a Gaussian with a mean metallicity of -1.57 and a dispersion of 0.57. Our smaller data set with a mean metallicity of -1.46 and dispersion of 0.55 is consistent with their sample; a one-sided KS-test shows that the smaller second peak at -0.61 is not statistically significant. Secondly, the {\elc} distribution, shown in Figure \ref{fig:dist}b, was first described by \citet{Wolfe2008}; they found that their sample of 37 positive detections of {\elc} showed a bimodal distribution. Our sample contains 32 positive detections, and shows a similar bimodality with peaks at $10^{-27.4}$ ergs s$^{-1}$ H$^{-1}$ and $10^{-26.7}$ ergs s$^{-1}$ H$^{-1}$ respectively. This allows the sample to be divided into two subsamples: a high-cool subsample, where {\elc} {\textgreater} $10^{-27}$ ergs s$^{-1}$ H$^{-1}$, and a low-cool subsample for which {\elc} {\textless} $10^{-27}$ ergs s$^{-1}$ H$^{-1}$. Finally, the {\w1526} distribution, discussed in \citet{Prochaska2008}, is shown in Figure \ref{fig:dist}c. While most DLAs exhibit saturated \ion{Si}{2}$\lambda$1526 lines, this distribution includes DLAs with unsaturated \ion{Si}{2}$\lambda$1526 lines to prevent a bias against low equivalent widths (see Section \ref{sec:bias}). The distribution is fully consistent with the sample found in \citet{Prochaska2008}. The fact that there exists good agreement between the sample considered in this paper and the larger samples used to study these three parameters in detail, shows that the current sample is a representative sample of DLAs for each of these three parameters. This is crucial for a proper exploration of correlations between the parameters.

The last parameter we consider is the {\delv} statistic. The distribution is shown in Figure \ref{fig:dist}d. Unlike previous papers \citep{Prochaska1997, Prochaska2008}, we do not use linear bins, but divide the bins logarithmically. This choice was made because for the current sample size, the velocity width distribution is well-approximated by a log-normal distribution. The {\delv} distribution has two important characteristics. The first important characteristic of this distribution is the sharp decline with decreasing {\delv} of DLAs with {\delv} below 17 {\kms} ($\log$ {\delv}$=1.23$). This decline is not likely to be an observational limit, because even our lowest resolution data has a {\delvres} of 8 {\kms} which corresponds to a {\delv} of 11 {\kms} for a single Gaussian component. Although some objects have been found with these small {\delv} values, they are extremely rare \citep{Cooke2011, Pettini2008, Penprase2010}. The rarity of these objects suggests that the majority of DLAs have at least one component with an internal velocity dispersion, $\sigma$, greater than 7 {\kms} or have multiple components which exhibit velocities of $\sim$ 10 {\kms} relative to each other. 

\input{./table4.tex}

The second characteristic of the distribution of velocity widths is the large number of high {\delv} DLAs. When plotted in a linear histogram, these DLAs form a `tail' in the distribution, as seen in \citet{Prochaska1997, Prochaska2008}. Most of these large {\delv} DLAs have {\delv} values below 200 {\kms}, but there are a few DLAs which show a velocity width significantly above this value \citep[e.g.][]{Petitjean2002}. These DLAs with very large velocity width are likely due to certain quasar sightlines encountering multiple galaxies or winds \citep{Prochaska2008}. As a result of the large number of high {\delv} DLAs, the resultant median {\delv} is significantly higher than expected from numerical simulations \citep{Prochaska1997, Razoumov2006, Pontzen2008}. The median value for our sample is  70$_{-13}^{+16}$ {\kms}. The uncertainty on the median is the combined uncertainty of three different factors. First, due to the finite sampling of the {\delv} distribution, our sample of 100 DLAs might not evenly be divided by the true median of the sample. To estimate the uncertainty due to this factor, we note that this is a simple selection problem which can be described by the binomial distribution with $p=0.5$. The binomial distribution has a 1-$\sigma$ uncertainty of $1/2\cdot\sqrt{n}$, where n is the number of data points.  This means that for a sample of 100 DLAs, there is a probability of 0.68 that the true median of {\delv} is contained in the range between DLAs 45 and 55 when the DLAs are arranged in order of increasing {\delv}. To this uncertainty, we need to add in quadrature the uncertainty arising from the limited sampling of the data. This uncertainty we take to be the average spacing of the $\sqrt{n}$ points centered around the median. Finally, we add in quadrature the mean observational uncertainty to get the uncertainty on the median. 

\subsection{Correlations and Dependencies}
\label{sec:cor}
In this section we discuss some of the two-parameter correlations and dependencies that exist between the six parameters discussed in this paper. Table \ref{tab:tst}  lists all 15 combinations of two parameters. For all 15 combinations of parameters, we are not only interested in any potential correlation, but we also want to know if the distribution of one parameter is dependent on the other. Therefore we apply a variety of different tests described below which test for both correlations and such dependencies.

To test for the existence of any potential correlation, we find a linear fit to the data using the $\tt MPFITEXY$ routine \citep{Williams2010}, which depends on the $\tt MPFIT$ package by \citet{Markwardt2009}. This routine takes into account the uncertainties in both parameters when calculating the slope and y-intercept of the best fit line. To calculate the uncertainty of the slope and y-intercept, we use a bootstrap method. The bootstrap method works by randomly selecting 100 DLAs from the original sample of 100 DLAs, but allowing for repeats. The resultant sample is then fitted using the same fitting routine as the original sample, and the slope and y-intercept are recorded. This sampling and fitting is repeated 1000 times, creating a distribution of slopes and y-intercepts. The 1-$\sigma$ uncertainty on the slope and y-intercept is inferred from a Gaussian fit to these distributions. A second test for the existence of a correlation between the parameters is provided by the application of a Kendall Tau test. The resultant two-sided significance of its deviation from zero is shown in Table \ref{tab:tst}; here a small value indicates a probable correlation.

To test if the distribution of one parameter is dependent on a second, we apportion our complete sample into three equally sized subsamples based on the value of the second parameter. We then compare the distribution of the first parameter for the subsamples with the smallest and largest values of the second parameter using four different tests. The first test we apply is a two-sided Kolmogorov-Smirnov (KS) test. This test will provide a probability that the two subsamples are drawn from the same parent population. A value smaller than 0.05 indicates that we can assume that the two populations were drawn from a different parent population at the 95\% confidence level (c.l.). To test if the variance of the two subsamples is significantly different, we use the F-statistic. Again a value of 0.05 or smaller indicates a significantly different variance in the two subsamples at the 95\% c.l. Finally, we used a Student's T-test and the Mann-Whitney U-test to compare the mean and median of the two subsamples, where again a smaller than 0.05 probability would indicate that the two subsamples have significantly different means or medians. The results of these tests for all 15 combinations are shown in Table \ref{tab:tst}. 

\begin{figure}[b]
\plotone{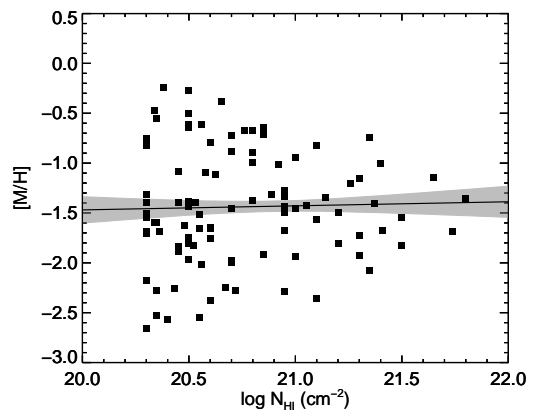}
\caption{Metallicity versus {\hi} column density. The square points mark our sample, and the solid line is a linear fit to the data, where the gray area marks the 1-$\sigma$ error on this line. The sample shows a lack of  high {\hi} column density - high metallicity systems. The lack of these systems has been noted in previous papers such as \citet{Boisse1998}. However, the sample also shows a lack of high {\hi} column density - low metallicity systems. Using the F-statistic, we show that the lack of these systems is not likely (i.e. less than 5\% probability) due to the small number of high {\hi} column density systems.}
\label{fig:mhvsnhi}
\end{figure}

In the next few subsections we will discuss some possible correlations that have not been discussed in previous papers; in particular the dependence of the {\hi} column density to the other parameters. We also explore the possible correlation between redshift and kinematics, which is important for exploring the interplay between redshift, metallicity and mass of DLAs. 

\begin{figure*}[t]
\epsscale{1.00}
\plottwo{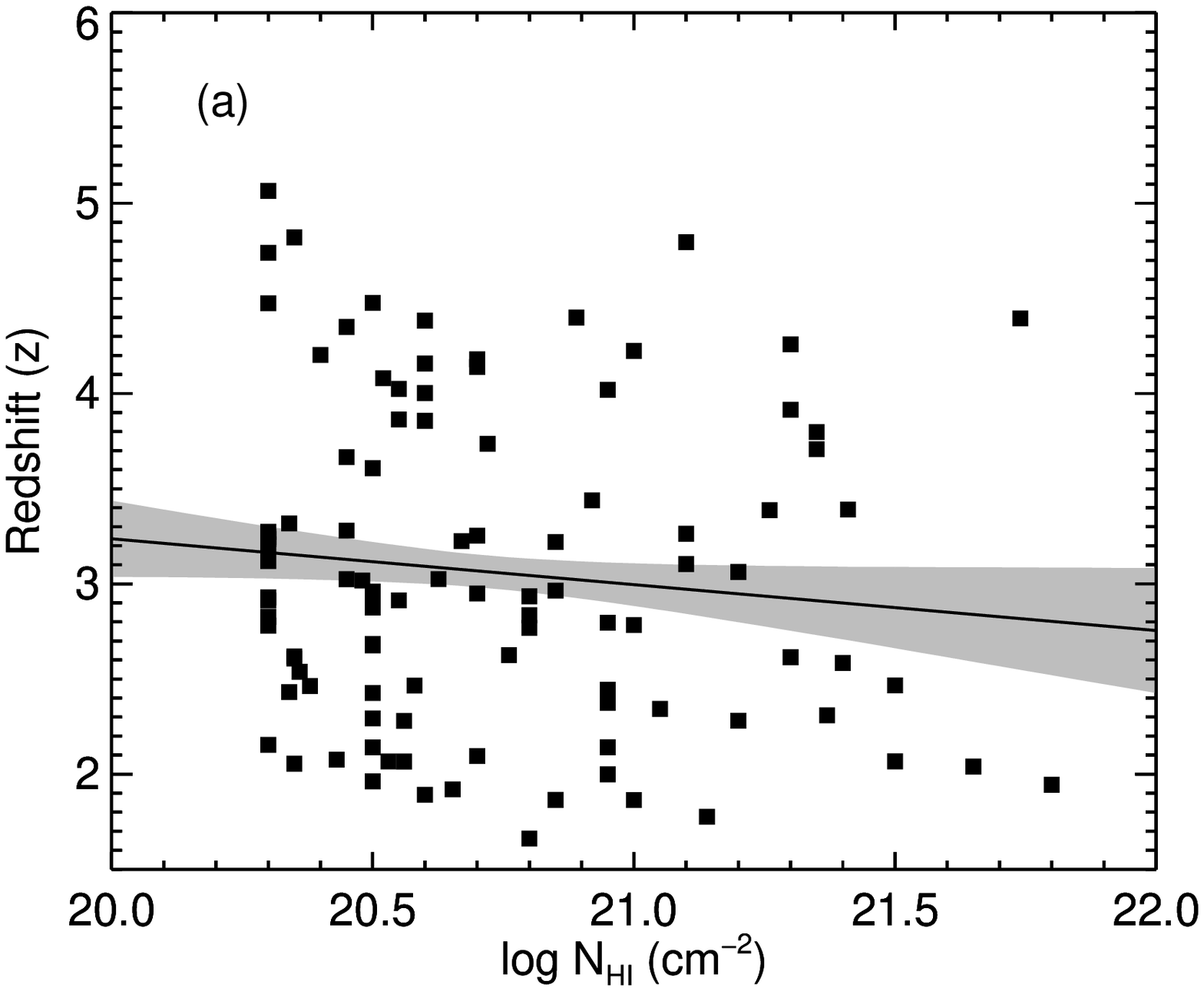}{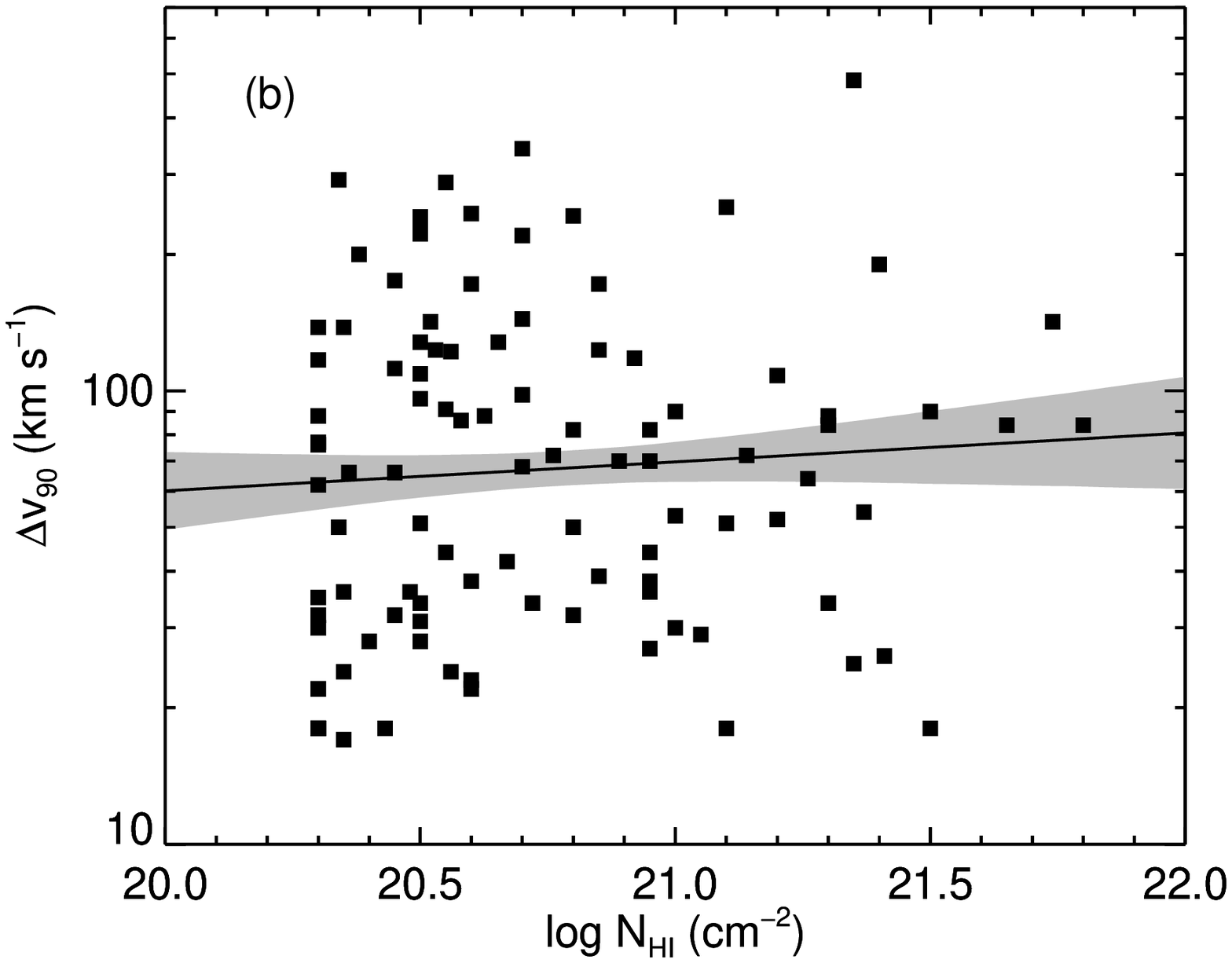}
\epsscale{1.00}
\plottwo{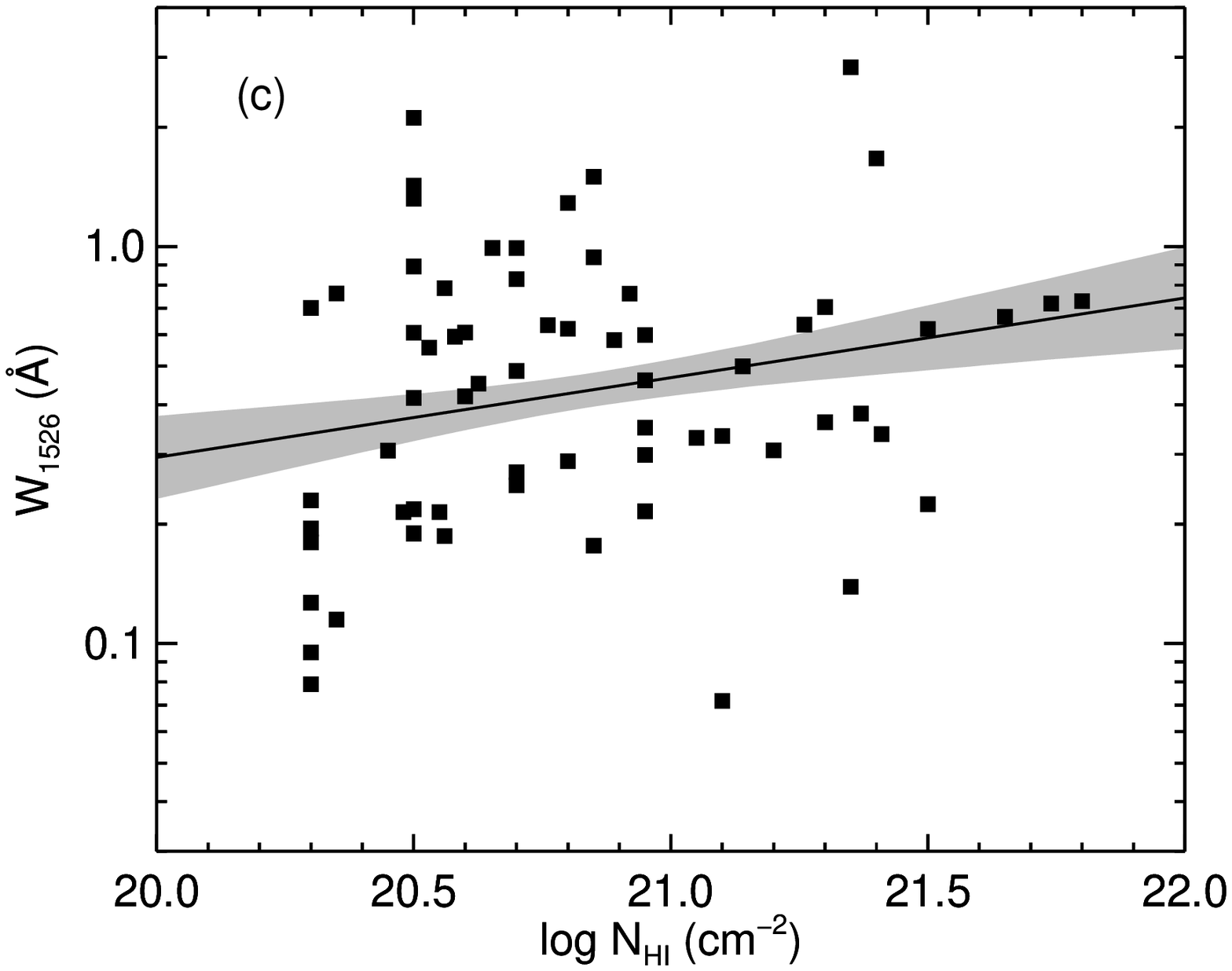}{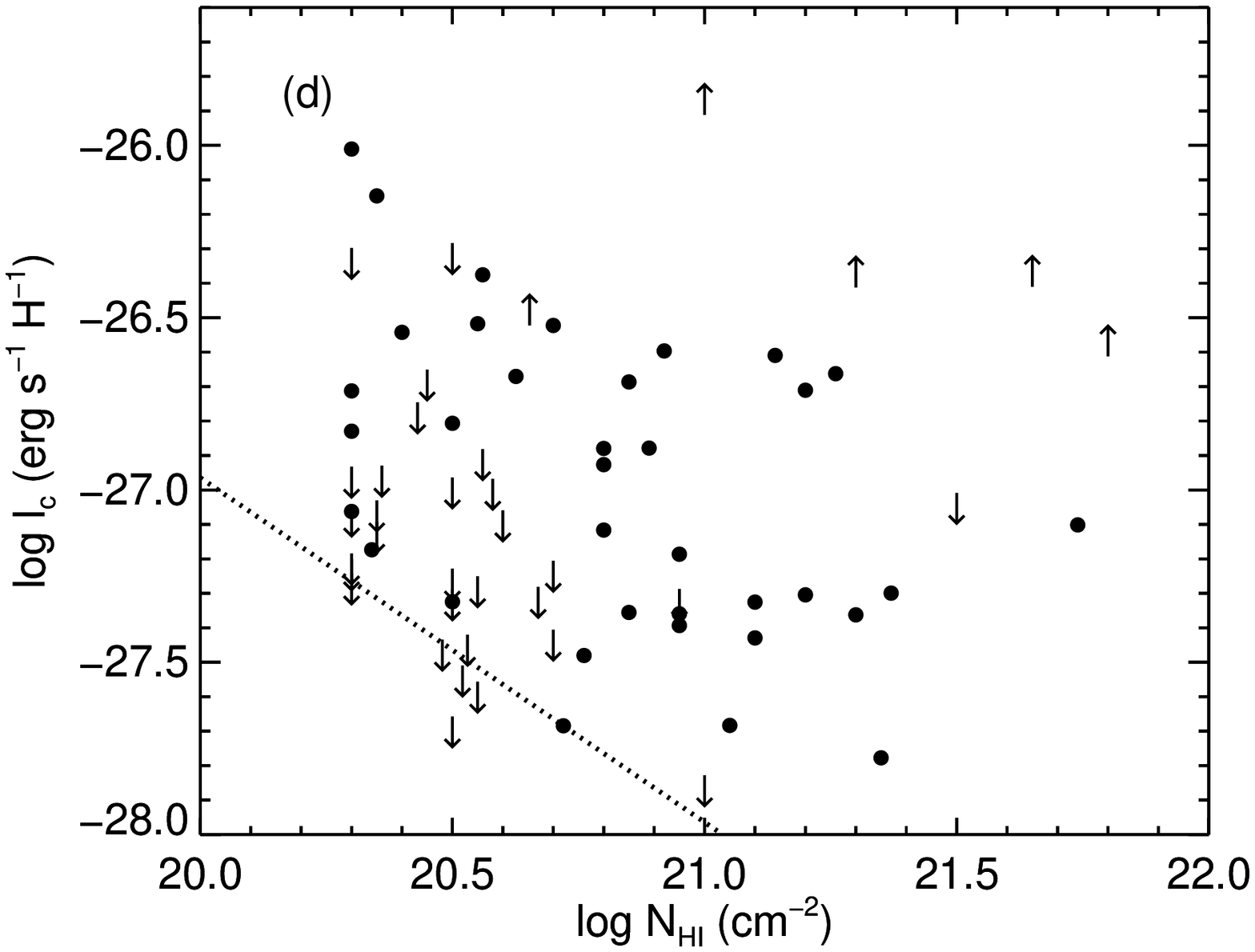}
\caption{Remaining four parameters plotted against the {\hi} column density. The figure markings are the same as Figure \ref{fig:mhvsnhi}. Both redshift and {\delv} are consistent with no correlation, whereas {\w1526} shows a potential correlation with the {\hi} column density, although this is not statistically significant for the sample size in this paper. The {\elc} distribution tentatively shows that the bimodality is more evident at higher column densities. The dotted line indicates the line below which we cannot reasonably positively detect a cooling rate, because of the required S/N needed for a positive detection.}
\label{fig:pvsnhi}
\end{figure*}

\subsubsection{[M/H] vs {\nh}}
\label{sec:mhnh}
Figure \ref{fig:mhvsnhi} shows a plot of metallicity versus {\hi} column density for the objects in our sample. The linear line was calculated using the method described in the previous section, and the shaded gray area marks the 1-$\sigma$ uncertainty on this line. There are two features of this plot we would like to discuss. First, the sample shows a lack of high {\hi} column density - high metallicity systems. \citet{Boisse1998} attributed this lack to a dust bias; presumably these systems would have large enough dust fractions to block out all of the light from the QSO and therefore escape detection. This assertion has been questioned by many papers including \citet{Ellison2001b, Ellison2005, Akerman2005, Jorgenson2006, Frank2010} who found that radio-selected DLAs do not differ significantly from those selected optically, and \citet{Kaplan2010} who detected certain metal-strong DLAs above the threshold found in \citet{Boisse1998}. A second interpretation for this lack was put forth by \citet{Schaye2001b}, which was further explored by \citet{Krumholz2009}, who showed that the absence of high {\hi} column density systems could be due to a transition from the atomic to molecular state of the atoms in the cold phase of a two-phase medium. However, both of these explanations only describe the lack of high column density - high metallicity systems, and are \emph{unable to explain the second feature, which is the lack of high {\hi} column density - low metallicity systems}. In fact, both features seem to be quite symmetric, in that the scatter plot exhibits a reflection symmetry about the line [M/H] $=-1.43$. To test if these features are statistically significant, we use the F-statistic. Table \ref{tab:tst} shows that the F-statistic for the [M/H] vs {\nh} correlation is 0.013. This indicates that the null hypothesis, which is that the smallest {\hi} column density DLAs (i.e. DLAs with  {\nh} $\leq 20.5$) and the largest {\hi} column density DLAs (i.e. DLAs with  {\nh} $\geq 20.85$) have the same variance, can be ruled out at a 95\% confidence level. Hence, both features are not likely due to small number statistics; possible explanations for the existence of these two features and the symmetry between them are given in Section \ref{sec:disccor}.

\begin{figure*}[t]
\epsscale{1.10}
\plottwo{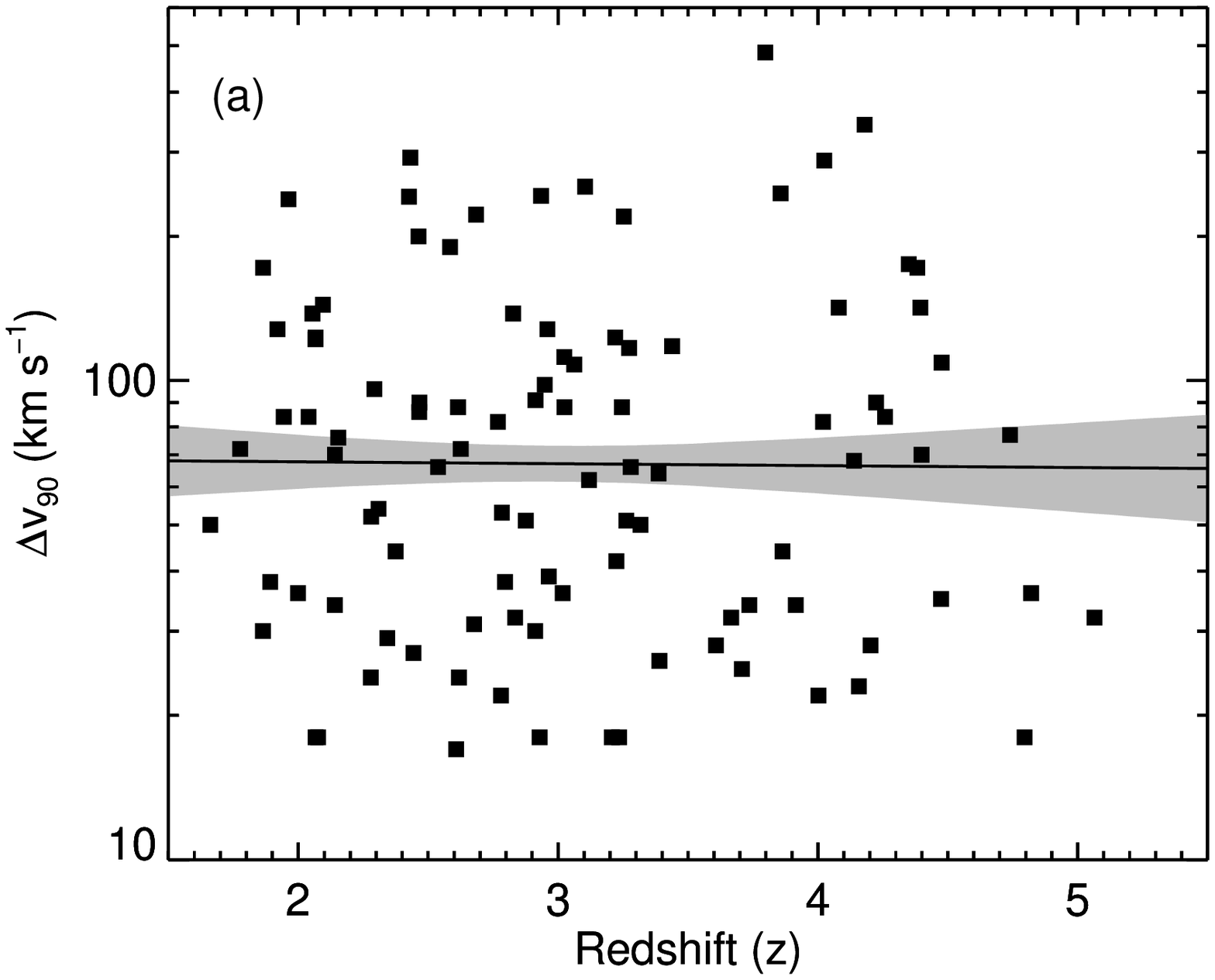}{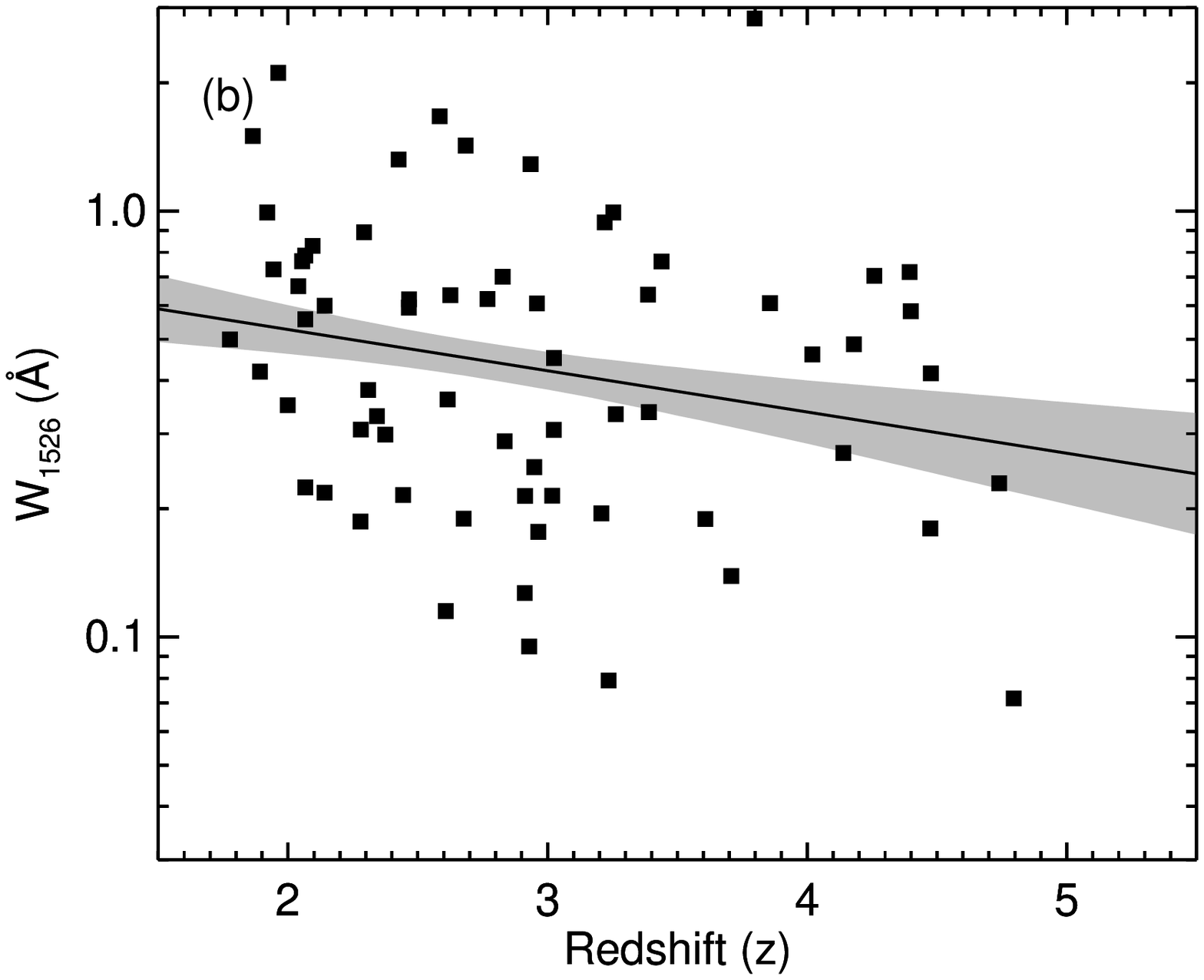}
\caption{Redshift evolution of the two different kinematic parameters. The {\delv} statistic shows no evidence for any redshift evolution. The linear fit is clearly consistent with a flat line. The {\w1526} statistic shows potential evidence for redshift evolution as can be seen by the solid linear regression line. However, the sample size  is too small to confirm this correlation.}
\label{fig:dvandwvsz}
\end{figure*}

\subsubsection{{\nh} dependencies of the remaining parameters}
\label{sec:pnhi}
Figure \ref{fig:pvsnhi} shows the remaining four parameters plotted vs {\hi} column density. Both redshift and {\delv} show no statistically significant correlation with {\nh}, and the distribution of these two parameters is also not dependent on {\nh}. For {\w1526} there is a greater likelihood that we see evidence for an increase in equivalent width with {\hi} column density as indicated by the Kendall tau parameter. However, this trend is not significant. This can be seen in the slope of the linear regression line which is consistent with no correlation at the 2-$\sigma$ level. Finally, for the {\elc} parameter, the high {\nh} subsample has a significantly smaller mean and median than the low {\nh} subsample. This is due to two effects. First, we can observe a similar trend as in the above case for metallicity that higher {\hi} column density systems tend to have less variance compared to lower {\hi} column density systems. Since the {\elc} distribution is bimodal, the higher {\hi} column density systems have values close to either of the two mean values, and therefore show a more distinct bimodality than the lower {\hi} column density systems. Secondly, we are unable to measure low {\hi} column density, low-cool systems, because we only use positive detections of the \ion{C}{2}$^*$$\lambda$1335.7 line for calculating {\elc}, and due to the limited signal to noise ratio of the spectra we are unable to positively detect the very low CII* column density systems. Both these effects contribute to the differing {\elc} distribution for the low {\nh} and high {\nh} subsamples. 

We would like to point out that this second effect also biases the relative sizes of the low-cool and high-cool subsamples. \citet{Wolfe2008} report relative sizes of the two subsamples to be 38\% for the low-cool subsample and 62\% for the high-cool sample using only positive detections. To gauge the extent of this bias, we estimate that for a spectrum with a signal to noise ratio of 30, the minimum value we can measure for the column density of CII* is approximately $10^{12.5}$ cm$^{-2}$. Using, equation \ref{eq:lc} we can compute the corresponding {\elc} value as a function of {\hi} column density. This boundary, below which we cannot make any positive detections, is shown in Figure \ref{fig:pvsnhi}d by a dotted line. This shows that the result quoted in \citet{Wolfe2008} underestimates the number of DLAs in the low-cool sample. Comparing the number of DLAs in the two subsamples for DLAs with {\hi} column density above 10$^{20.7}$cm$^{-2}$, for which this bias is negligible, shows that there are in fact more DLAs in the low-cool subsample than the high-cool subsample. However, we caution that not enough data points are available to make a precise estimate of the sizes of the two subsamples.

\subsubsection{{\delv} and {\w1526} vs redshift}
\label{sec:dvandwvsz}
The last two possible correlations that we consider are the two kinematic properties, {\delv} and {\w1526}, as a function of redshift. The kinematical properties are believed to be strongly linked to the mass of the dark matter halo. This is a ubiquitous feature of almost all DLA models \citep[e.g.][]{Haehnelt1998, Prochaska1997}. Since current $\Lambda$CDM models predict hierarchical galaxy formation, which in turn predict an increase in the mass of dark matter halos over time, we may expect to see a corresponding increase in the {\delv} and {\w1526} statistic of the DLAs as well. Again we use the tests described in Section \ref{sec:cor} to see if there is any linear trend in the data, and/or if the distribution of {\delv} or {\w1526} is evolving with redshift. 

Figure \ref{fig:dvandwvsz}a shows {\delv} as a function of redshift. As in Figure \ref{fig:mhvsnhi}, the solid black line indicates the best fit linear line to the data, and the gray area marks the 1-$\sigma$ uncertainty on the parameters. The resultant best fit is: log$_{10} ({\Delta}\rm{v_{90}}) = (0.00{\pm}0.04) {\cdot} z+(1.84{\pm}0.13)$. This linear regression line is clearly consistent with no redshift evolution. Furthermore, all other tests show that the distribution of {\delv} is also consistent with no evolution.

This is in contrast with a previous result found by \citet{Ledoux2006}, who found that the mean {\delv} decreased with increasing redshift when they apportioned their sample into two subsamples based on redshift. When we apportion our sample into two subsamples based on redshift, we find two main differences between their results and ours. First of all, our sample covers a larger redshift range; in particular our median values for each redshift subsample are 2.400 and 3.722 whereas their median redshifts are 2.087 and 2.796; any redshift evolution should therefore be more noticeable in our subsample. Second of all, their sample was smaller and no errors were reported in the median value of the velocity widths. Indeed, if we calculate the errors on the median for their two subsamples, we get a median of 69$^{\rm{+25}}_{\rm{-13}}${\kms} and 89$^{\rm{+20}}_{\rm{-13}}$ for the high redshift and low redshift subsample respectively. This shows that the two medians are within 1-$\sigma$ of each other, indicating that their sample is also consistent with no redshift evolution.

\begin{figure*}[t]
\epsscale{1.7}
\plottwo{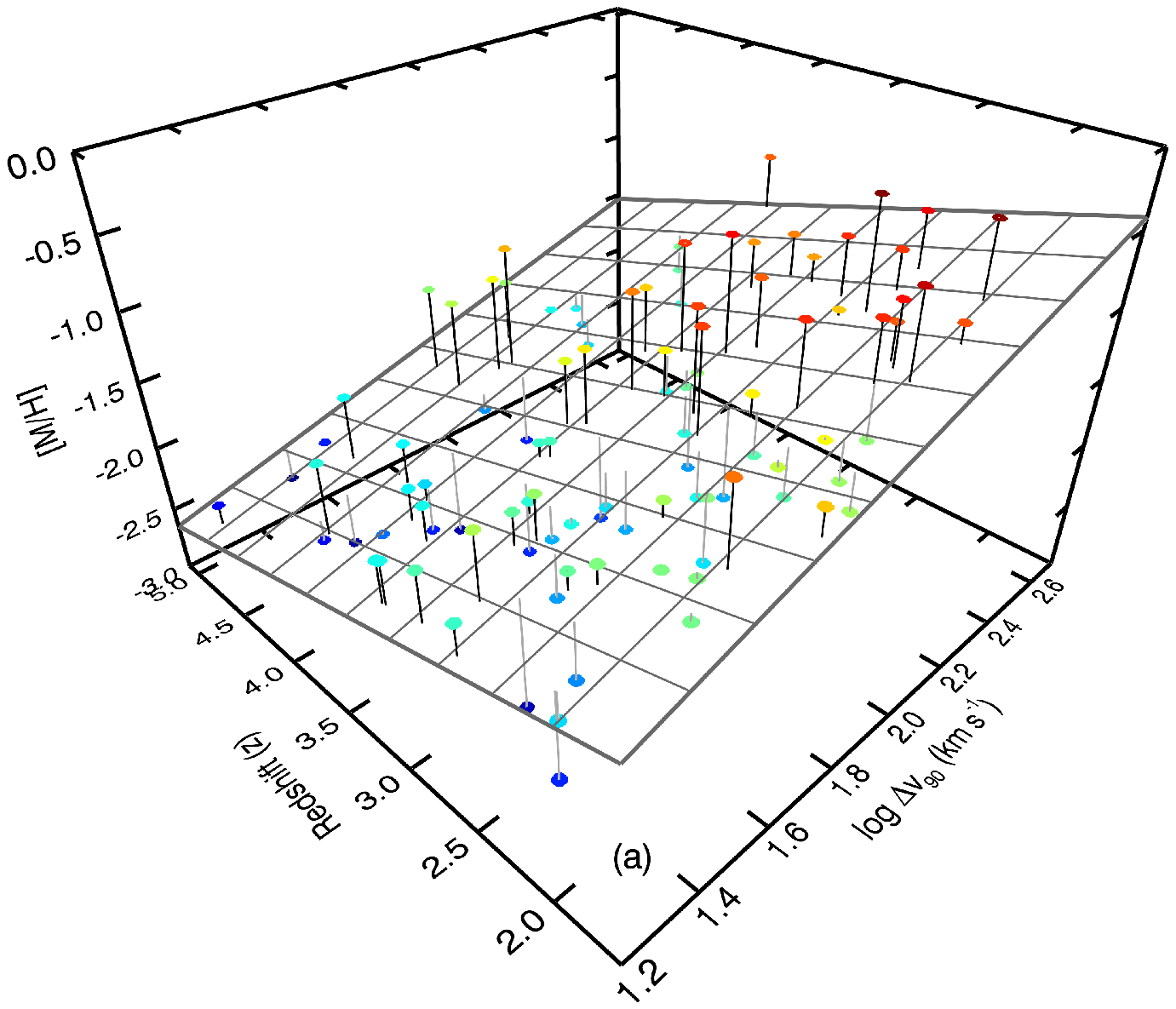}{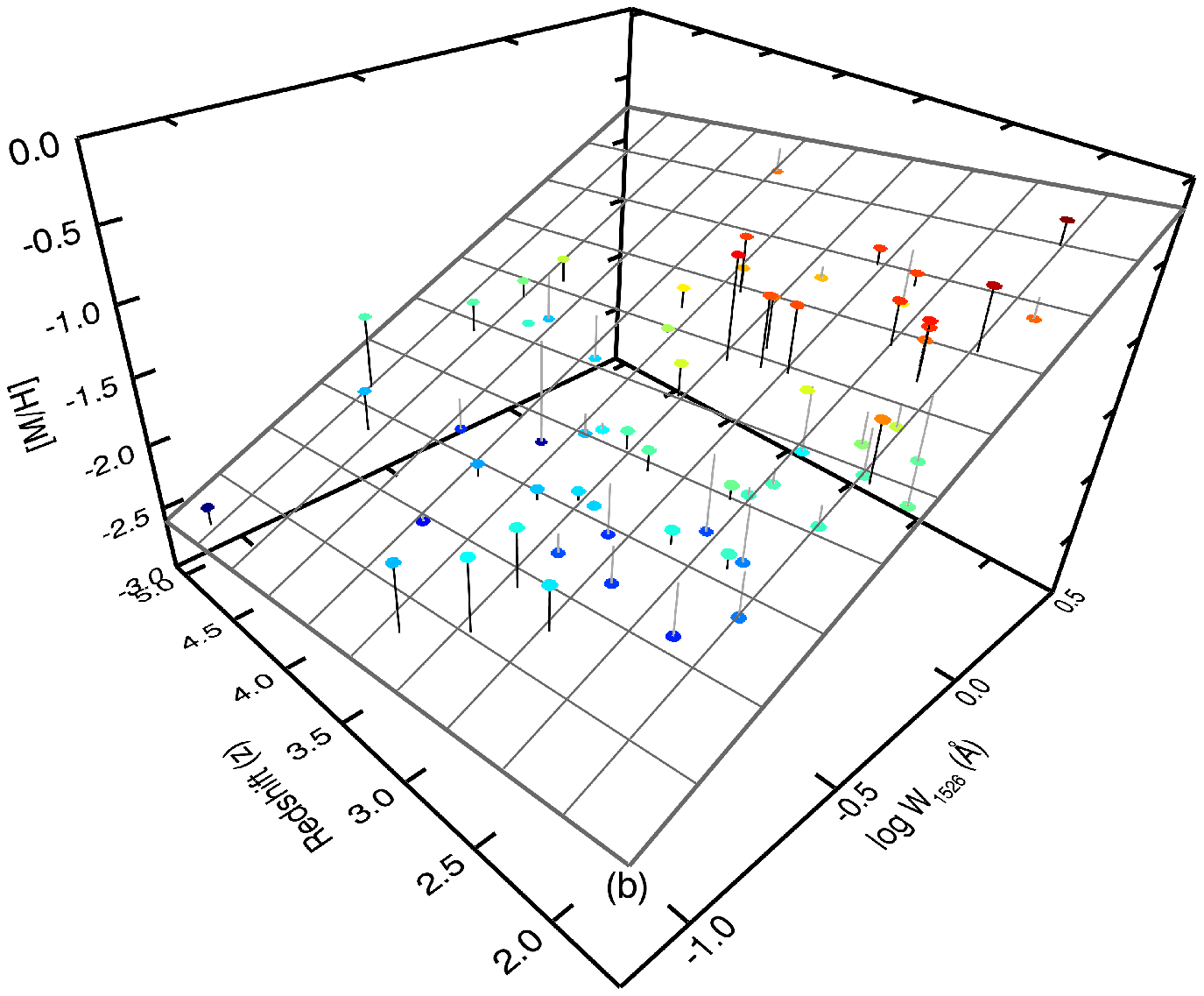}
\caption{Panel (a) of this figure illustrates that when redshift, {\delv} and metallicity of a DLA are plotted on a three dimensional scatter plot, the points trace a plane in this parameter space, although with scatter. This plane is also seen when the {\delv} parameter is replaced by {\w1526} (panel (b)). The plane is marked by thick, dark gray lines. The data points are connected to this plane by solid black (gray) lines if they fall above (below) the plane. The grayscale (color scale) of the data points is correlated to the metallicity of the DLA.}
\label{fig:fund0}
\end{figure*}

\begin{figure*}[t]
\epsscale{0.9}
\plottwo{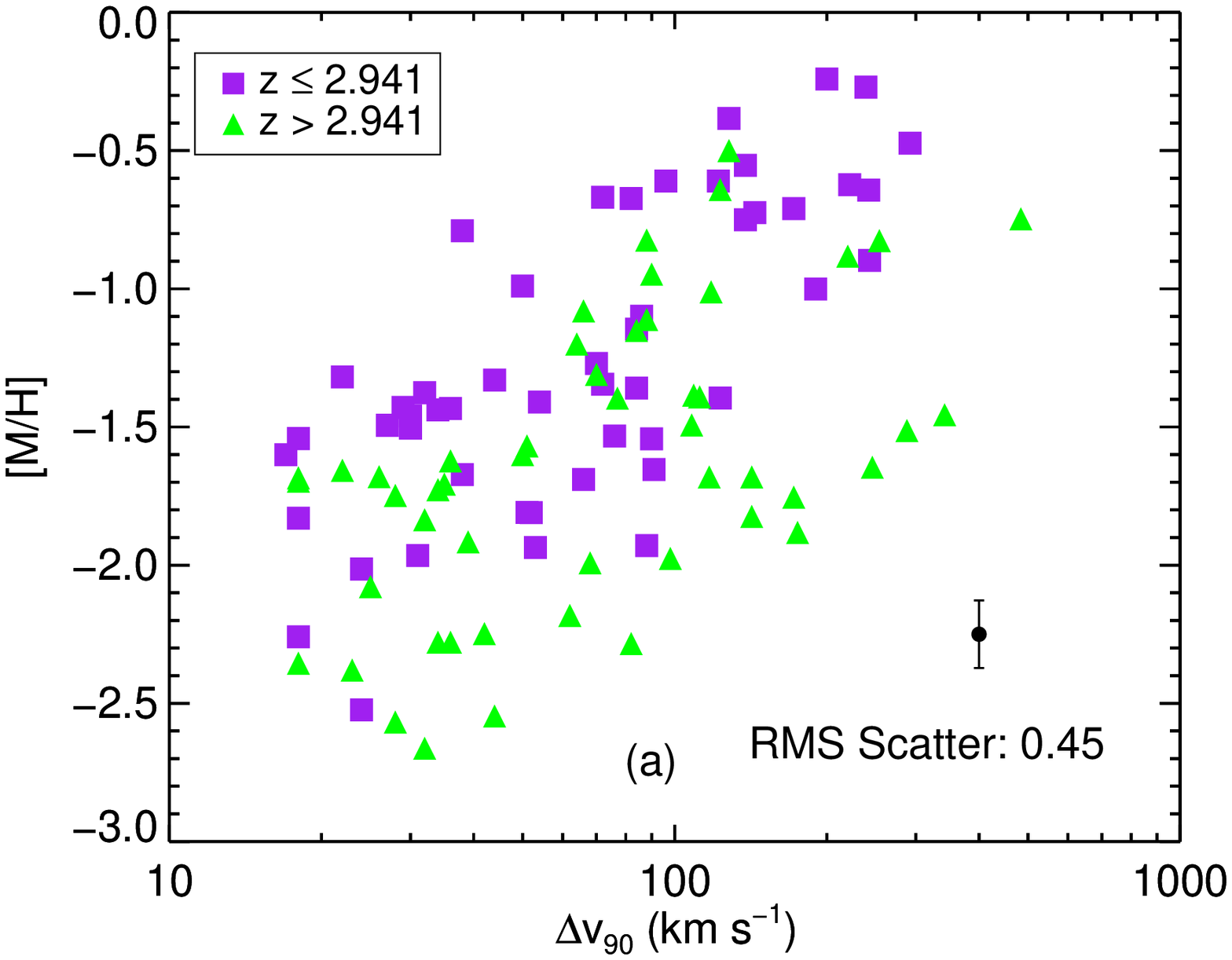}{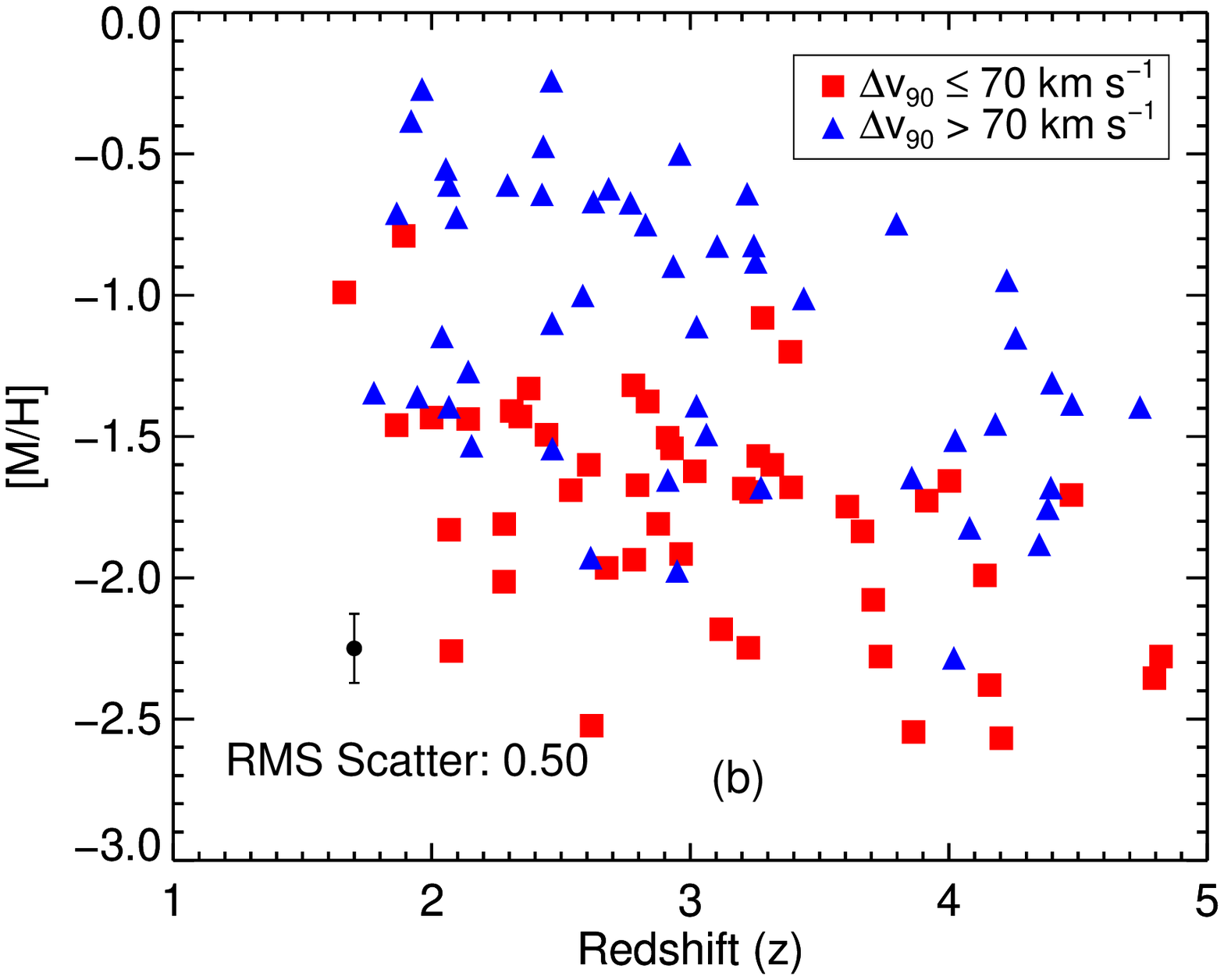}
\plottwo{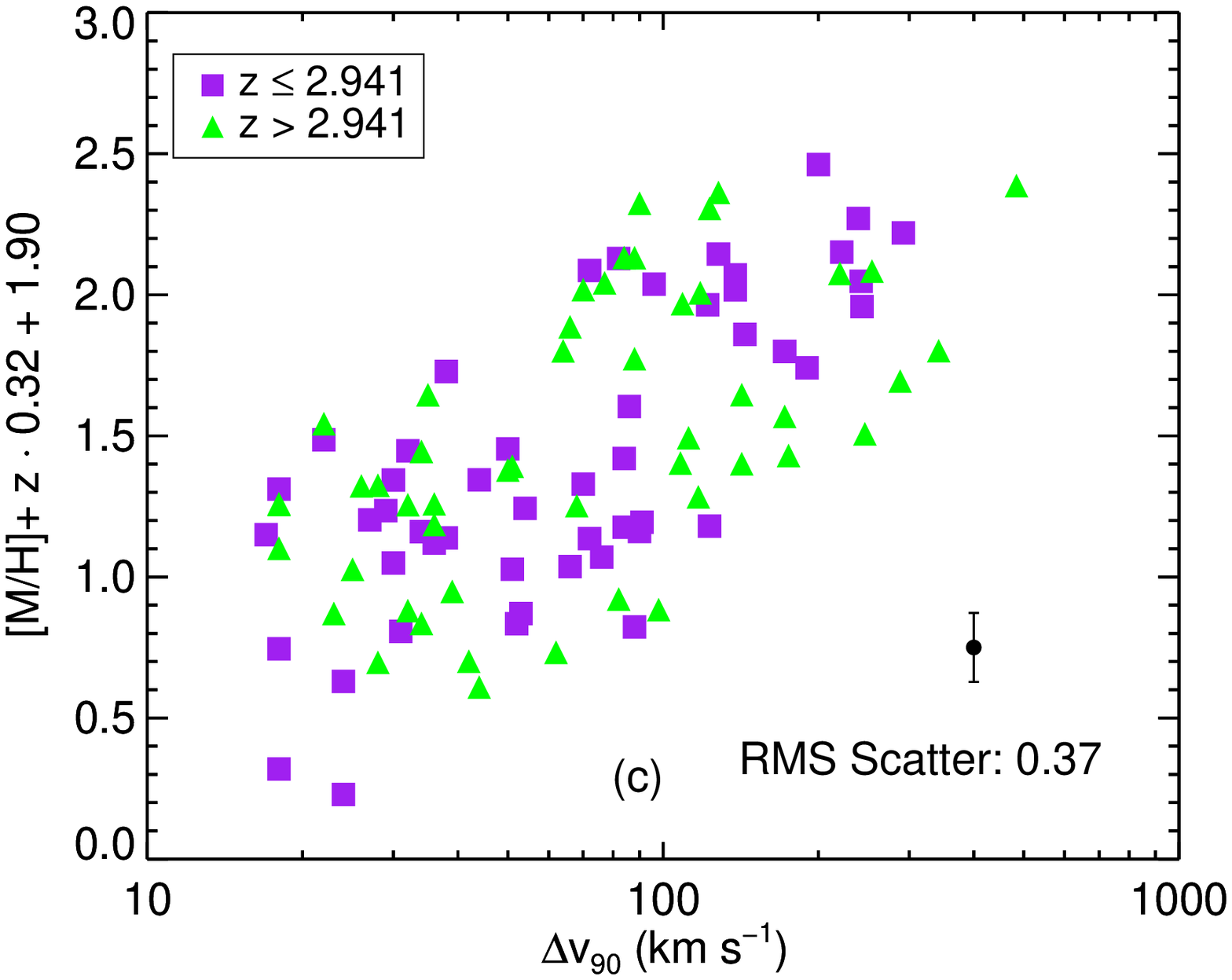}{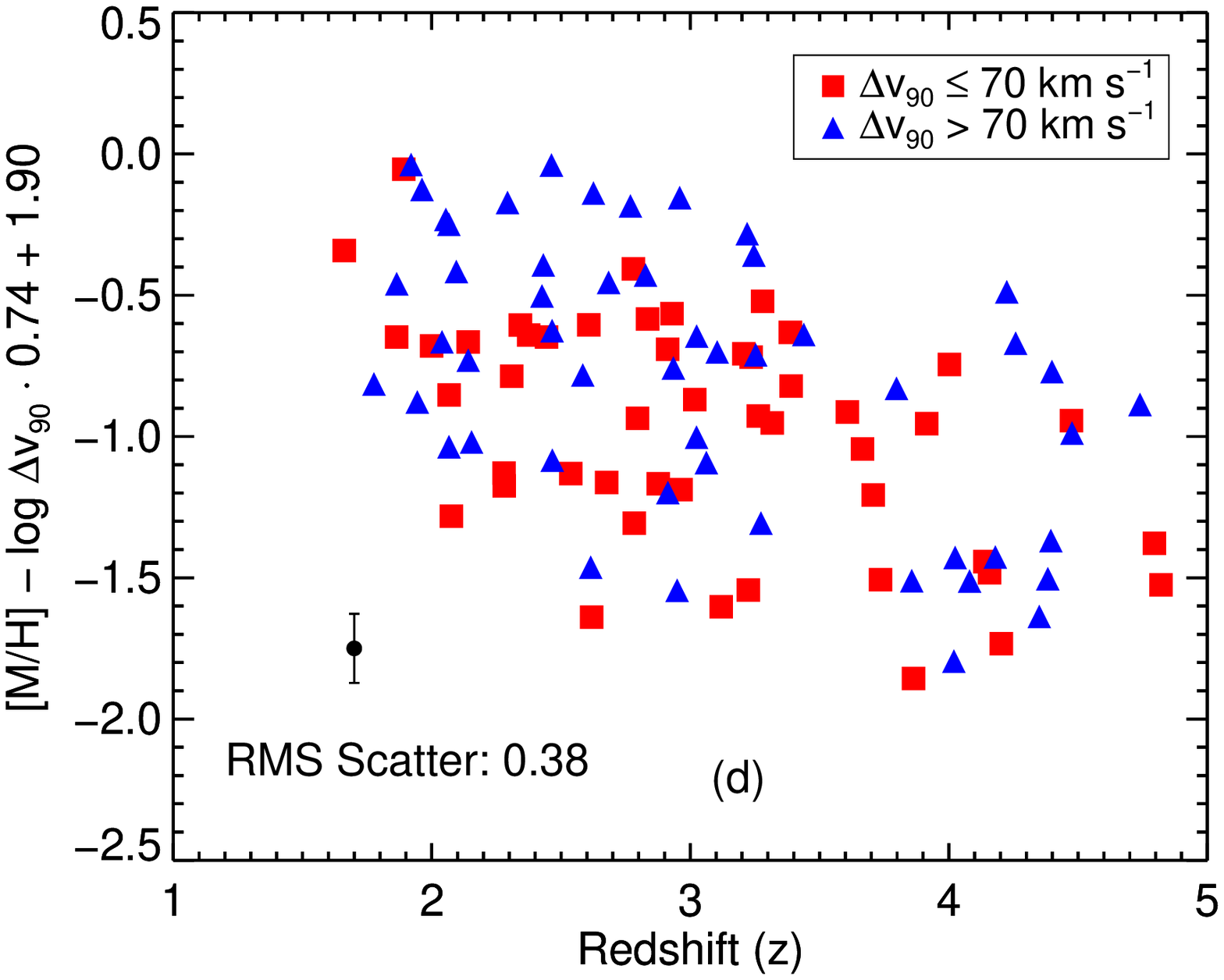}
\caption{This figure illustrates the advantage of using the fundamental plane equation in describing the two correlations shown in the top two figures. Panel (a) shows the metallicity-{\delv} correlation where we have apportioned our sample into two equal subsamples based on redshift. The solid (green) triangles are the higher redshift DLAs, whereas the solid (purple) squares are lower redshift DLAs. This panel clearly shows that a large part of the scatter is due to the correlation between redshift and metallicity. Similarly, the scatter in the metallicity-redshift correlation (panel (b)) is due to the existence of a mass-metallicity correlation at each redshift. By applying the fundamental plane equation we can reduce this scatter as is shown in panels (c) and (d). The scatter is still bigger than the mean observational uncertainty, which is shown by the errorbar.}
\label{fig:fund}
\end{figure*}

The second kinematic parameter we consider is {\w1526}, shown in Figure \ref{fig:dvandwvsz}b. The data shows a lack of high-redshift large-equivalent width systems, and the Kendall Tau correlation test shows a small value of 0.042 indicating that a potential correlation could exist. However, a linear fit to the data gives log$_{10}(W_{1526})=(-0.10{\pm}0.05) \cdot z - (0.08{\pm}0.15)$, which indicates that the correlation is seen at only the 2-$\sigma$ level. Moreover, when we compare the median and mean of the highest redshift DLAs with those of the lowest redshift DLAs, we do not see a significant difference in their value. Indeed the KS test shows a likelihood of 56\% that the two samples are drawn from the same parent population. We therefore conclude that with the current sample size, the correlation between {\w1526} and redshift cannot be determined at $>$ 3-$\sigma$ significance level.

\section{The Fundamental Plane(s) of DLAs}
\label{sec:fund}
As was discussed in the introduction, the aim of this paper is to examine the interplay between multiple parameters of DLAs using the fundamental plane description. For this description to be useful, we need to have three parameters where two parameters are not strongly correlated, but each parameter shows a strong correlation to the third parameter. In Table \ref{tab:tst} we list all of the correlations that exist between the six parameters discussed in this paper. The only parameters that clearly satisfy this criterion are metallicity, redshift and {\delv}, and metallicity, redshift and {\w1526}. When we plot the first three parameters for each DLA on a three-dimensional scatter plot, we see that the points indeed fall close to a plane inside this space, although with scatter (Figure \ref{fig:fund0}a). Figures \ref{fig:fund}a and \ref{fig:fund}b are two projections of this plane along the redshift and {\delv} axis respectively, where the third parameter is apportioned into two equally sized subgroups. Figure \ref{fig:fund}a is similar to Figure 2 of \citet{Ledoux2006} and Figure \ref{fig:fund}b is similar to Figure 11 of \citet{Rafelski2012}. Figure \ref{fig:fund}a shows that the scatter in the {\delv}-metallicity correlation is in part due to redshift evolution of metallicity and similarly the scatter in the redshift-metallicity correlation is in part due to differences in the kinematics of the DLAs at each redshift.

\begin{figure*}
\epsscale{0.9}
\plottwo{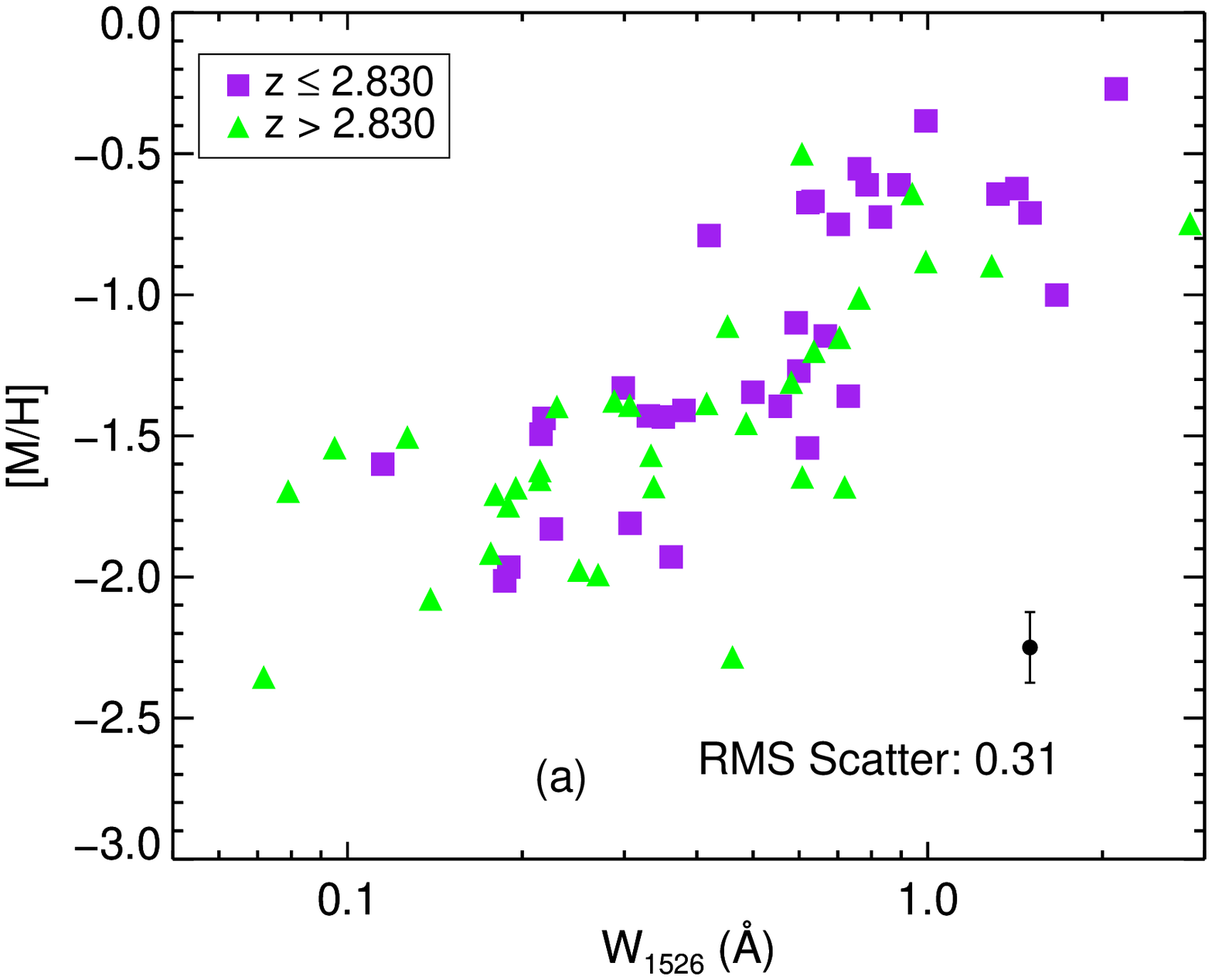}{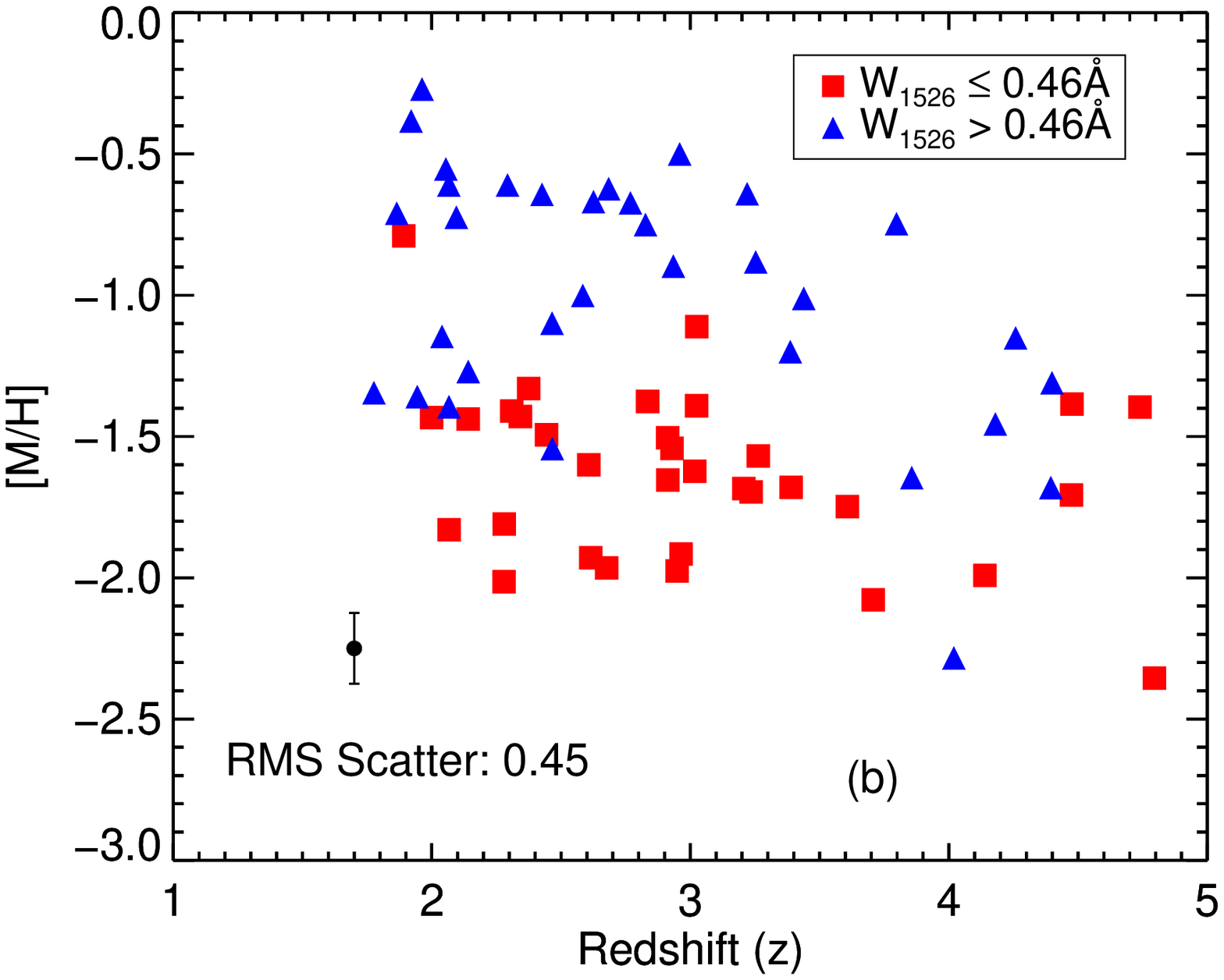}
\plottwo{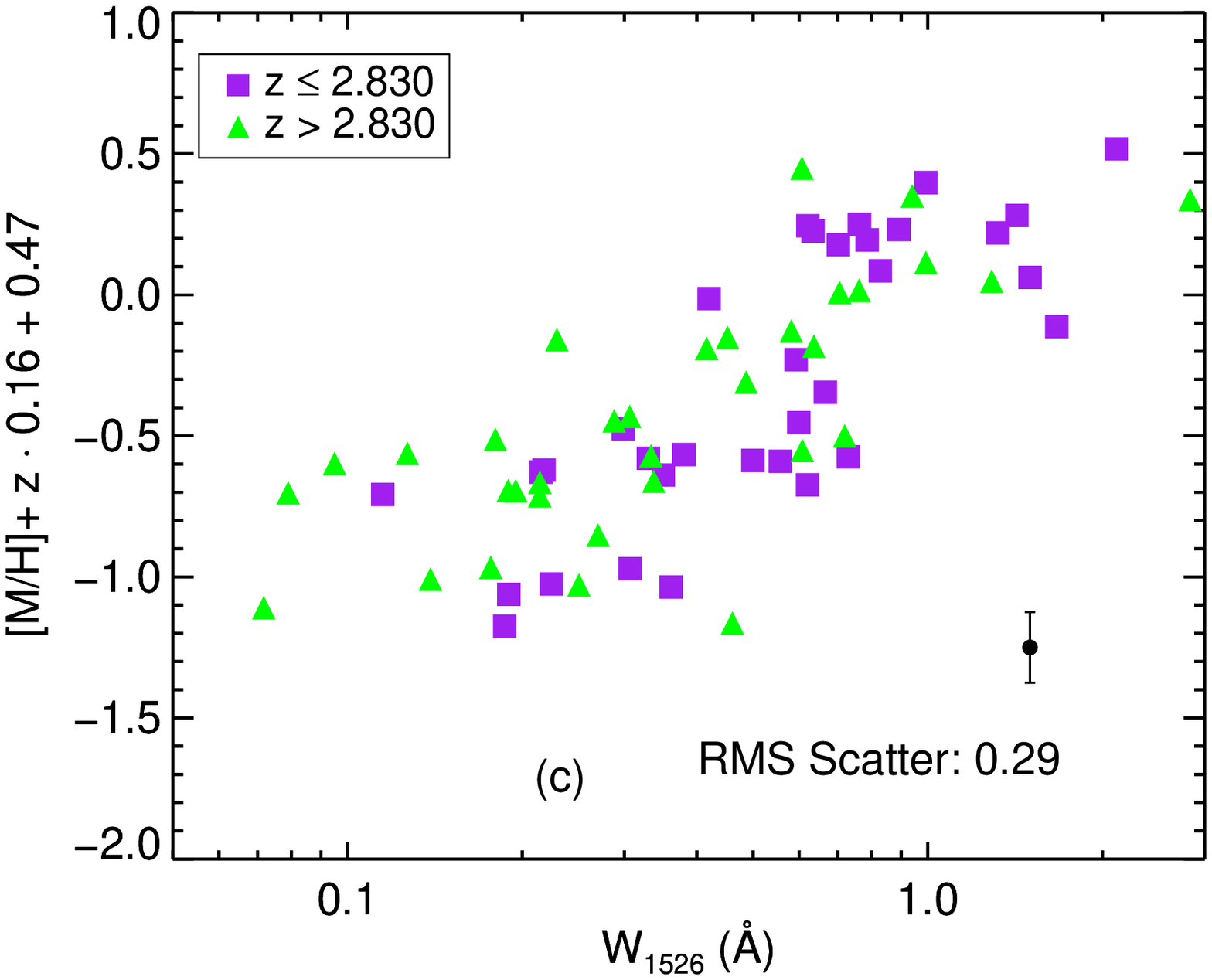}{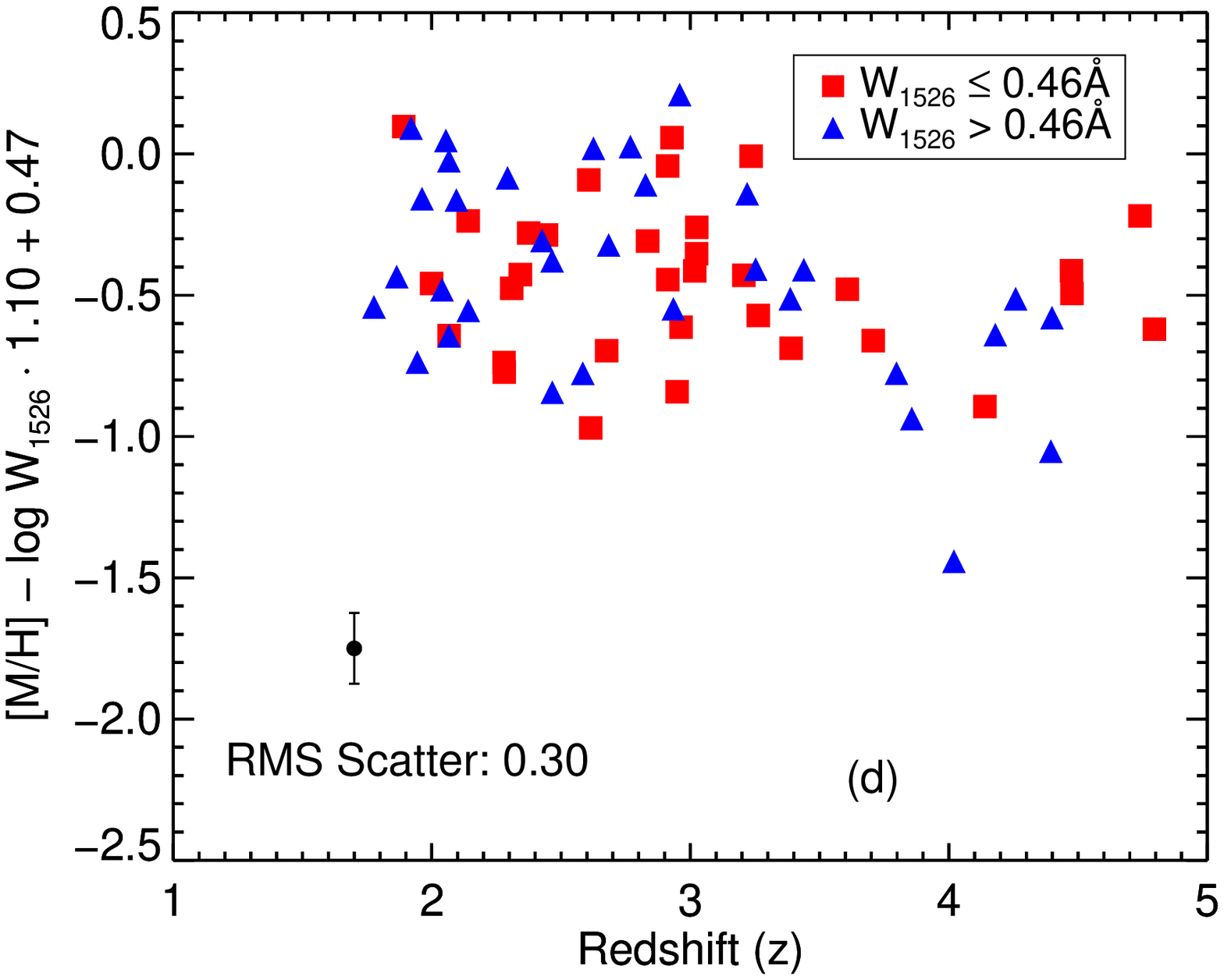}
\caption{This figure illustrates the fundamental plane using the {\w1526} statistic instead of {\delv}. The panels are the same as Figure \ref{fig:fund}, except the kinematic parameter has been switched. Note that the scatter in the equivalent width-metallicity relationship is not significantly reduced by applying the fundamental plane equation. This is due to the fact that the plane is only slightly tilted in the redshift direction. This, in part, explains the observational result that the equivalent width-metallicity relationship exhibits the smallest scatter of any parameter with metallicity \citep{Prochaska2008}.}
\label{fig:fund2}
\end{figure*}

To accurately describe the shape of the distribution in this three-dimensional space, we need to fit a plane equation to the scatter plot. To do this, we use the `direct fit' least square fitting technique described in \citet{Bernardi2003},  which fits a fundamental plane equation of the form $\rm{[M/H]}=a{\cdot}{\log}{\Delta}v_{90}+b{\cdot}z+c$ to the data set. To account for measurement errors, we weigh each individual point by the inverse of the estimated uncertainty using the methods described in \citet{Sheth2012}. This produces the following covariance matrix, $R$ which is normalized by the RMS value of each quantity:
\[
\bordermatrix{~&\text{\footnotesize{${\Delta}$v$_{90}$}}&\text{\footnotesize{$z$}}&\text{\footnotesize{[M/H]}} \cr
\text{\footnotesize{${\Delta}$v$_{90}$}}&1.000 & 0.121 & 0.361 \cr
\text{\footnotesize{$z$}}&0.121 & 1.000 & -0.396 \cr
\text{\footnotesize{[M/H]}}&0.361 & -0.396 & 1.000 \cr}
\]

From this matrix we can calculate the coefficients of the plane using the equations described in \citet{Bernardi2003}. The resultant plane is described by the following equation:
\begin{equation}
\begin{split}
\rm{[M/H]}=(-1.9{\pm}0.5)+(0.74{\pm}0.21){\cdot}{\log}{\Delta}v_{90}\\
-(0.32{\pm}0.06){\cdot}z
\end{split}
\label{mhdv-eq}
\end{equation}
The uncertainty of the parameters are estimated using the bootstrap method discussed in Section \ref{sec:cor}. It is important to note that because {\delv} and redshift are not correlated (Figure \ref{fig:dvandwvsz}a), the uncertainties in the coefficients of the fundamental plane are independent of each other. Figure \ref{fig:fund}c and \ref{fig:fund}d are plots in which we take into account the third parameter using the plane equation described above (i.e. this is like rotating the plane until it is edge-on). To quantify the reduction in scatter, we calculate the RMS scatter around a linear fit to the scatter plot before and after we apply the plane. The scatter in the correlations before we apply the plane equation is 0.45 dex (0.50 dex) for the metallicity-{\delv} (redshift-metallicity) relationship. After we apply the plane the scatter is reduced by approximately 20 \% to 0.37 dex (0.38 dex) for each of the correlations. The reduction in scatter is clearly visible in Figures \ref{fig:fund}c and \ref{fig:fund}d. However, the scatter is still significantly bigger than the observational uncertainty on each measurement, which is on average 0.12 dex. We will comment more on the reduction in scatter in Section \ref{sec:discfund}.

The second fundamental plane of interest is similar to the first, except the {\delv} statistic is replaced by the other kinematic parameter, {\w1526} (Figure \ref{fig:fund0}b). \citet{Prochaska2008} showed that the metallicity-{\w1526} correlation exhibits smaller scatter than the metallicity-{\delv}  relationship, and a fundamental plane equation between metallicity, redshift and {\w1526} may reduce this scatter further. As in Sections \ref{sec:pnhi} and \ref{sec:dvandwvsz}, we are using just the subset of {\w1526} measurements with saturated {\w1526} values, although the inclusion of these systems does not significantly affect the shape or tilt of the fundamental plane. The equation of the fundamental plane between redshift, metallicity and {\w1526} is:
\begin{equation}
\begin{split}
\rm{[M/H]}=(-0.47{\pm}0.14)+(1.1{\pm}0.3){\cdot}{\log}W_{1526}\\
-(0.16{\pm}0.06){\cdot}z
\end{split}
\label{mhew-eq}
\end{equation}

Figure \ref{fig:fund2}a shows metallicity versus {\w1526}; unlike Figure \ref{fig:fund}a where the {\delv}-metallicity trend moves downward with increasing redshift (i.e. at higher redshift, a given {\delv} corresponds to a lower metallicity), the {\w1526}-metallicity correlation does not evolve with redshift. The lower redshifts are higher on the correlation trend line because of the correlation between redshift and metallicity, but very little of the scatter is due to redshift, and therefore a fundamental plane description does not reduce the scatter significantly (Figure \ref{fig:fund2}c) To be specific, the scatter is reduced by only 0.02 dex. Visually this means that we are already looking edge-on to the plane in Figure \ref{fig:fund2}a. However, this second fundamental plane is able to reduce the scatter significantly (0.15 dex) in the redshift-metallicity relation (Figure \ref{fig:fund2}d) as was the case with the previous fundamental plane. The measured scatter around the two correlations after applying this fundamental plane is 0.29 dex for the {\w1526}-metallicity correlation and 0.30 dex for the redshift-metallicity correlation. The scatter around this plane is significantly smaller than the scatter around the fundamental plane found previously, which we discuss further in Section \ref{sec:discfund}.

\section{Discussion}
\label{sec:disc}
The primary aim of this paper has been to explore the fundamental relations that exist between the parameters that describe DLAs. In this section we discuss the results of this study.

\subsection{Implications of two parameter dependencies}
\label{sec:disccor}
Table \ref{tab:tst} lists all of the possible two parameter dependencies for DLAs that are explored in this paper. The first notable dependency is that between the {\hi} column density and the metallicity of the DLA (Section \ref{sec:mhnh}). Previous work \citep[e.g.][]{Boisse1998} observed a lack of high {\hi} column density, high metallicity systems. Our sample, which was selected in a less biased way, shows that this lack of high {\hi} column density, high metallicity systems is just half of the picture, and there is also a lack of high {\hi} column density low metallicity systems. In Section \ref{sec:mhnh} we have given two possible explanations for the lack of high {\hi} column density, high metallicity systems, but we need an explanation for the lack of high {\hi} column density, low metallicity systems. 

One possibility is that this lower `envelope' is not related to the upper `envelope'. The existence of just a lower envelope can be explained by the following two explanations: (1) higher {\hi} column density DLAs are due to sightlines probing the inner part of galaxies, because of the metallicity gradients in DLAs \citep{Chen2005}, these sightlines should increase in metallicity. (2) Low metallicity DLAs are due to sightlines crossing cold flows, which because of their low density and high ionization level are unable to produce sightlines with a high {\hi} column density. The latter is seen in simulations modeling cold flows by \citet{Fumagalli2011} who find that only at higher redshifts cold flows are able to host low {\hi} column density, low metallicity DLAs.

The main problem with both these explanations is that they do not explain the striking symmetry that exists between the upper and lower envelope, and instead attribute the symmetry to coincidence. Two possible explanations that instead take the symmetry as a premise are (1) the fundamental plane equation is {\hi} column density dependent. If indeed the tilt of the plane is slightly different for higher column density systems (i.e. their metallicity does not evolve with redshift as much and/or their mass-metallicity correlation has a slightly different slope), then this could reduce the scatter in their metallicity. Our current data set is lacking enough high {\hi} column density DLAs to confirm or refute this explanation. (2) Higher {\hi} column density systems contain more absorption components than lower {\hi} column density systems. The thought behind this is that each dark matter halo consists of components with varying metallicities. A low {\hi} column density sightline through this halo will only sample a few components, and therefore can experience a wide range in metallicities. On the other hand, a high {\hi} column density sightline will sample many components, and its metallicity will be the average metallicity of all these components. As a result the scatter in metallicities for these large {\hi} column density systems is smaller than for the low {\hi} column density systems. By varying the average metallicities of each dark matter halo depending on redshift and mass, we can still reproduce all other correlations with this explanation. Both explanations are therefore fully consistent with current observations.

\begin{figure}[t]
\plotone{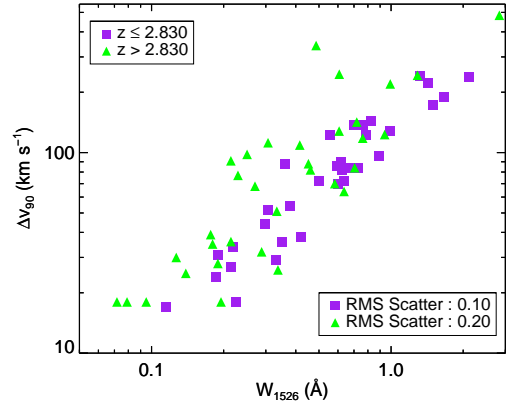}
\caption{Plot of the two kinematic properties where the sample has been apportioned into two subgroups based on the redshift of the DLA. The low redshift subgroup, solid (purple) squares, shows a small scatter of 0.10 dex around a linear fit to the data, whereas the higher redshift sample, solid (green) triangles, shows a significantly larger scatter of 0.20 dex. This increase in scatter is only toward higher {\delv} values for a given {\w1526} value. In the text we provide two possible explanations for this observation.}
\label{fig:dvvsew}
\end{figure}

Besides the dependency between metallicity and {\hi} column density, the second set of notable dependencies discussed in this paper are those between redshift and the kinematic parameters, {\delv} and {\w1526}. Both kinematic parameters are believed to trace the mass of the dark matter halo that hosts the DLA, although with significant scatter \citep[e.g.][see further Section \ref{sec:discfund}]{Ledoux2006, Prochaska2008}. Therefore, these two parameters should show a strong correlation as is displayed in Figure \ref{fig:dvvsew}. In this figure we have apportioned the sample into a low redshift and high redshift subsample. Interestingly, the high redshift sample has twice the RMS scatter compared to the low redshift subsample. Moreover, the scatter at high redshift only increases upward of the {\delv}-{\w1526} trendline at low redshift, suggesting that for a given {\w1526} value, the maximum available velocity width, {\delvmax}, increases with increasing redshift. Since the scatter around the {\w1526}-metallicity correlation is smaller than the scatter around the {\delv}-metallicity correlation, \citet{Prochaska2008} argues that {\w1526} is a better indicator of the dark matter halo mass than {\delv}. Under this assumption, the above result suggests that for a specific dark matter halo mass (i.e. a specific value of {\w1526}), {\delvmax} increases with redshift. This result, however, cannot be statistically confirmed with the current sample size.

Nevertheless, we provide here two potential explanations for the increase in scatter with increasing redshift around the {\delv}-{\w1526} correlation depending on which model is used to describe DLAs. If we assume that DLAs are large massive disks \citep[e.g][]{Prochaska1997}, then one way to interpret this result is that for a given dark matter halo mass, the rotational velocity structure of the neutral gas increases with increasing redshift, which will increase {\delvmax}. This is in agreement with the theoretical model of \citet{Mo1998}, which predicts that for a given dark matter halo mass the rotational velocity of the galactic disk will increase with increasing redshift. If instead we assume that DLAs are a collection of low-mass protogalactic clumps \citep[e.g][]{Haehnelt1998}, as in current $\Lambda$CDM models, we can interpret the result as a redistribution of the neutral gas content in dark matter halos with redshift. In this case, the {\delvmax} for a specific dark matter halo mass at high redshift is larger, because some sightlines will probe satellite galaxies which move in the potential of the primary galaxy increasing the relative velocity of these objects. At low redshift, more satellite galaxies will have merged, decreasing the probability of these sightlines, and therefore decreasing {\delvmax}. Together with a decrease in galactic scale winds at low redshift, \citet{Cen2012} has shown with numerical simulations that these two effects can decrease the {\delvmax} with decreasing redshift.

Although we see tentative evidence for an evolution in the scatter around the correlation between {\delv} and {\w1526} with redshift, no correlation is found between redshift and {\delv} as shown in Figure \ref{fig:dvandwvsz}a. Interestingly, this constancy in {\delv} with redshift is also seen in the numerical simulations by \citet{Cen2012}. \citet{Cen2012} attributes this 
constancy to two countering processes: the growth of the dark matter halo mass with decreasing redshift, and the decrease in galactic scale winds due to reduced star formation at lower redshifts. Unlike {\delv}, {\w1526} shows a potential correlation with redshift. However, this correlation cannot be confirmed above the 3-$\sigma$ level with the current sample size. Using a variety of different tests, we also looked at any potential evolution in the distributions of the kinematical parameters. Both distributions are consistent with no evolution. The lack of evolution of both the mean and distribution for both parameters is significant for two reasons. First, the lack of evolution in the distribution of either kinematic parameter indicates that at each redshift, DLAs are embedded in a wide range of dark matter halo masses. Second, the weak correlation between redshift and kinematics allows us to combine two separate correlations, namely the redshift-metallicity correlation and the mass-metallicity correlation, into a single fundamental plane equation.

\subsection{Implications of the fundamental plane of DLAs}
\label{sec:discfund}
The main result in this paper is the existence of a fundamental plane for DLAs. In Section \ref{sec:fund} we show that the redshift-metallicity and metallicity-velocity width correlations can be combined into a single planar equation, $\rm{[M/H]}=(-1.9{\pm}0.5)+(0.74{\pm}0.21){\cdot}{\log}{\Delta}v_{90}-(0.32{\pm}0.06){\cdot}z$. If instead we use the {\w1526} statistic, the equation becomes $\rm{[M/H]}=(-0.47{\pm}0.14)+(1.1{\pm}0.3){\cdot}{\log}W_{1526}-(0.16{\pm}0.06){\cdot}z$. The underlying reason for the existence of both planes is thought to be similar. Since both kinematic parameters are tracers of mass, the fundamental plane simply combines the redshift-metallicity correlation and the mass-metallicity correlation, which are correlations known to exist for DLAs, into a single equation. 

This equation provides a better description of the fundamental relation that exists between redshift, metallicity and kinematics than the two correlations for two reasons. First, it reduces the scatter around the correlations, and therefore provides a more stringent constraint for simulations modeling DLAs. The reduction in scatter is about 20\%, giving a reduced scatter of 0.37 dex and 0.38 dex around the correlations. This is still significantly larger than the observational uncertainty for each measurement, which is about 0.12 dex. However, we do not expect the scatter to reduce to the observational uncertainty because a single dark matter halo can produce a range in metallicities, velocity widths, and equivalent widths depending on where the quasar sightline intersects the dark matter halo. Hence, a certain amount of scatter is inherently part of the quasar absorption line experiment. To estimate the size of this scatter, we turn to numerical simulations such as those presented by \citet{Pontzen2008}. These simulations suggest a wide range of uncertainties depending on which model is used to describe DLAs, showing that a scatter of 0.38 dex can be reproduced from just intersecting dark matter halos at differing impact parameters. However, most models predict a scatter slightly smaller than this. The smaller scatter in the {\w1526} plane strengthens the idea that this parameter traces the kinematics of the halo gas \citep{Prochaska2008}, since the halo gas is assumed to be more spherically distributed than the neutral ISM, and therefore will be less affected by the impact parameter and inclination angle of the quasar sightline.

\begin{figure*}[t]
\plottwo{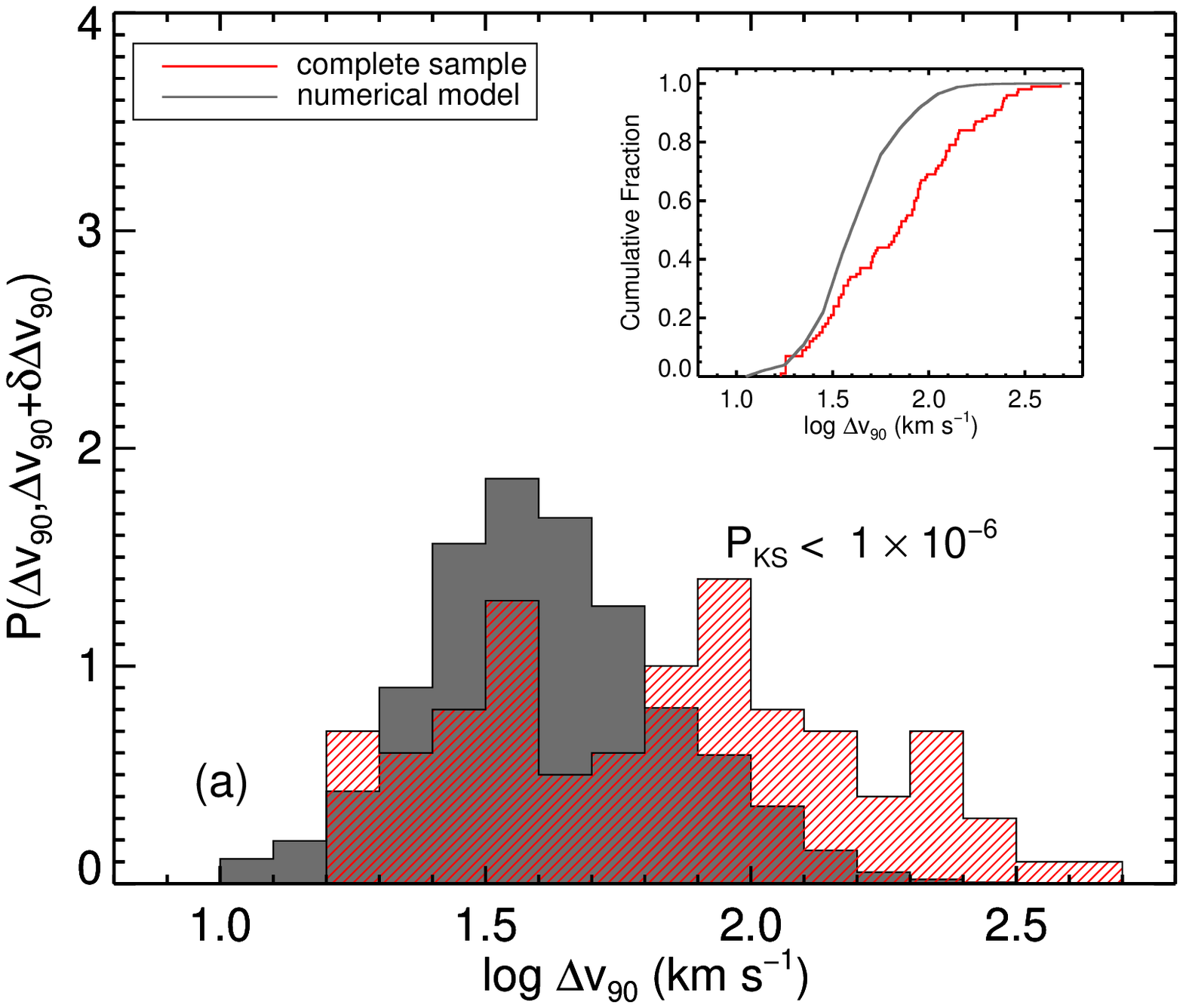}{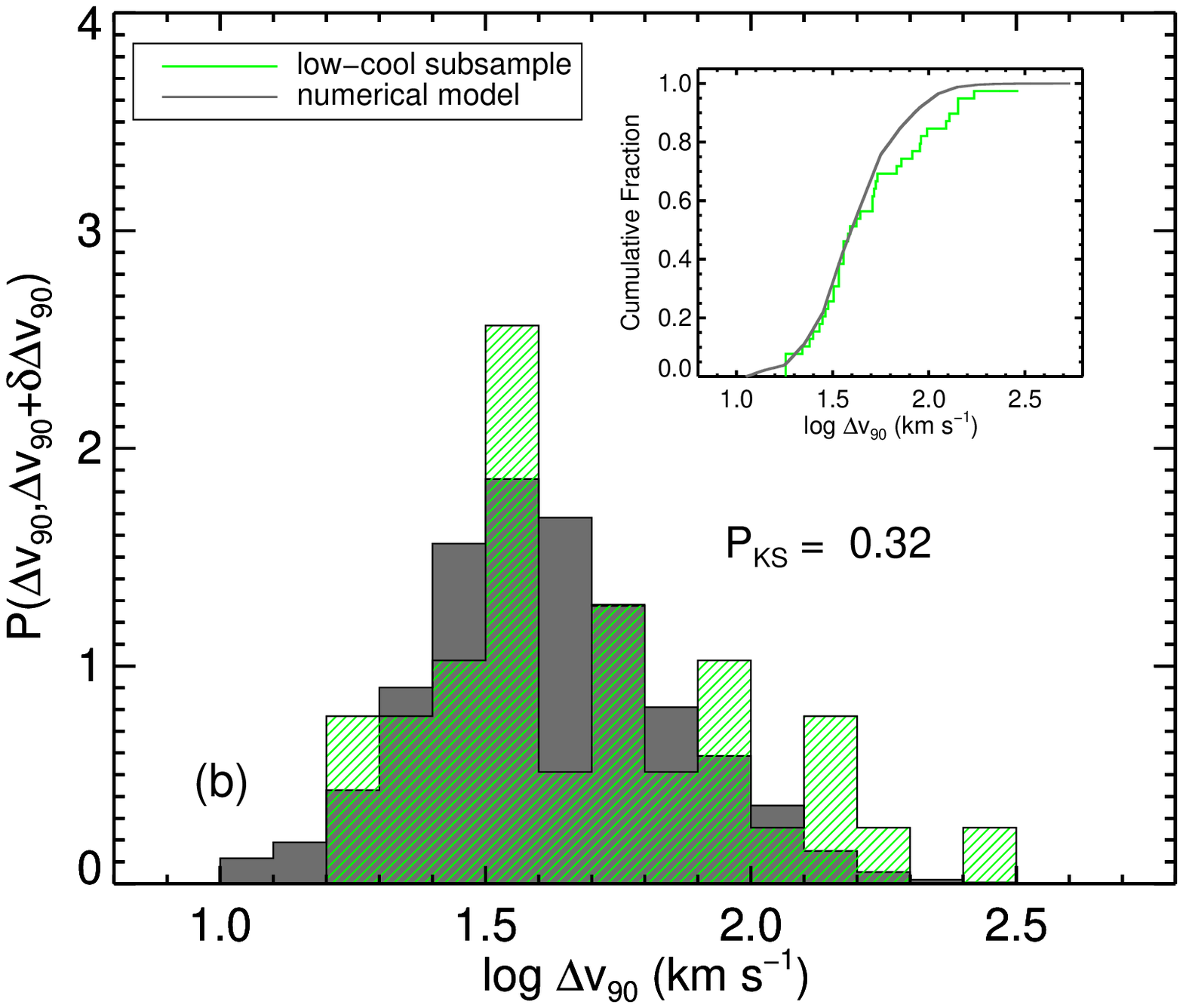}
\caption{Distribution of the kinematic parameter {\delv} for both the observational data and the numerical model by \citet{Pontzen2008}. The solid gray distribution is the distribution of the numerical model by \citet{Pontzen2008}. The hashed (red) sample in panel (a) is the complete sample, whereas the hashed (green) sample in panel (b) is just the subset of DLAs that are part of the low-cool population of DLAs as defined in \citep{Wolfe2008}. The left panel shows that the complete sample is inconsistent with the numerical model, since a two-sided KS test gives a probability less than 1 $\times$ 10$^{-6}$ that the two populations are drawn from the same parent distribution. On the other hand, the right panel shows that the low-cool subsample of DLAs is consistent with the velocity distribution found from the numerical model. Since low-cool DLAs are believed to arise from less massive dark matter halos, this result corroborates the hypothesis that numerical simulations are underproducing the occurrences of DLAs that reside in massive dark matter halos.}
\label{fig:dvmoddist}
\end{figure*}

The second reason the fundamental plane description is a better description than the two individual correlations is that its existence confirms that the dark matter halos hosting DLAs satisfy a mass-metallicity correlation \emph{at each redshift}. \citet{Ledoux2006} showed that this held true for their sample apportioned into a low redshift and high redshift subsample. Our result expands on their result and indicates that the mass-metallicity correlation holds over a larger redshift range. Moreover, we show that the zero point of the {\delv}-metallicty correlation is evolving with redshift. Figures \ref{fig:fund}a and \ref{fig:fund2}a show that both subgroups of DLAs apportioned by redshift follow a similar mass-metallicity correlation. This can also be seen in Figures \ref{fig:fund}b and \ref{fig:fund2}b where the more massive dark matter halos (i.e. larger {\delv} and larger {\w1526}) have at each redshift higher metallicities than the less massive dark matter halos. These two figures clearly show that at each redshift, the majority of the scatter in the metallicity-redshift correlation is caused by the existence of a mass-metallicity correlation. This result is consistent with the existence of mass-metallicity correlations found at both low and high redshift in star-forming galaxies \citep{Tremonti2004, Savaglio2005, Erb2006, Maiolino2008}. Although these studies used stellar mass and not total dynamical mass, there is a strong correlation expected between them \citep{Brinchmann2000}. 

We can use this dataset to explore the evolution of the mass-metallicity correlation for DLAs over the redshift range $z=2$ to $z=5$. At redshifts below 0.75, the stellar mass-metallicity correlation evolves with redshift in a mass-independent way; the slope of the correlation stays the same, but the zero point of the correlation decreases with increasing redshift \citep{Moustakas2012}. Interestingly, this decrease in zero point and constancy of the slope is also seen in Figure \ref{fig:fund}a for the {\delv}-metallicity correlation. Moreover, the decrease in zero point for the our DLAs is about 0.3 dex per unit redshift, which is comparable to the decrease in the mass-metallicity relationship seen at redshifts below 0.75 \citep{Moustakas2012} and between 0.7 and 3.5 \citep{Maiolino2008} for mass-metallicity correlations determined from stellar masses. This suggests that DLAs are enriched in a similar manner as the star-forming galaxies used to determine the stellar mass-metallicity correlation. In a recent paper, \citet{Moller2013} conclude that the zero point of the mass-metallicity correlation might not evolve past a redshift of 2.6. This is in contrast to our result, which shows that the zero point is steadily evolving over the redshift range z=2 to z=5. Currently the sample of unbiased DLAs is too small to distinguish one scenario from the other. On the other hand, the zero point of the {\w1526}-metallicity correlation is not evolving with redshift. This suggests that the {\delv} parameter is a better indicator of the total stellar mass inside the dark matter halo compared to the {\w1526} parameter. This is not surprising since {\delv} describes the kinematics of the neutral gas, which should be more strongly correlated to the stellar mass than the mass of the dark matter halo.

\subsection{Comparison with Models}
\label{sec:discmod}

We conclude this paper by comparing the fundamental plane equation with numerical models which have specifically modeled DLAs. To date, DLA simulations have tried to reproduce some of the distributions of parameters such as metallicity and {\delv} \citep{Pontzen2008, Tescari2009}, and the observed two-parameter correlations \citep{Pontzen2008, Cen2012}, but this paper introduces a more restrictive criterion which should be satisfied, the fundamental plane of DLAs. A successful simulation should be able to populate the fundamental plane in a similar manner to what is observed. However, currently no simulation tracks these three parameters accurately. Instead we therefore consider only projections of the fundamental plane. In particular, the projection along the {\delv} axis gives the metallicity-redshift correlation; its agreement with models has been discussed in detail in \citet{Rafelski2012}. The projection along the redshift axis leads to the metallicity-{\delv} correlation. This correlation has been compared to simulations in both \citet{Tescari2009} and \citet{Pontzen2008}, where the latter finds a correlation equation of: log$_{10}${\delv}$=2.5+0.58{\cdot}\rm{[M/H]}$ at redshift 3. They compared this to the observational correlation equation of \citet{Ledoux2006} and noted the good agreement except for the significantly larger observational scatter. Comparing the model of \citet{Pontzen2008} to the new fundamental plane equation at redshift 3 (i.e. setting $z=3$ in the fundamental plane equation) gives a similar agreement. The main improvement here, however, is in the scatter around this relationship. The scatter in previous observations was too large, because these observational analyses did not take into account the redshift evolution of metallicity. Using the fundamental plane reduces the scatter by about 20\%, and therefore the scatter in the \citet{Pontzen2008} simulation is in better agreement with the observational data then previously assumed.

Finally, we also compare the distributions of each of the parameters. Of these, the {\delv} distribution is the most interesting, because of the long standing inability of models in reproducing this distribution \citep{Prochaska1997, Haehnelt1998, Razoumov2006, Pontzen2008}. As was discussed in the introduction and shown in Section \ref{sec:dist}, this distribution has two characteristics that are difficult to reproduce by simulations based on $\Lambda$CDM models; namely the large median velocity, and the high-velocity tail of the distribution. It is believed that the inability of reproducing these features of the {\delv} distribution is due to the inability of the numerical simulations to produce enough DLAs that reside in massive dark matter halos compared to less massive dark matter halos \citep{Pontzen2008, Tescari2009}. We test this hypothesis by comparing the numerical simulations to just those DLAs that are believed to reside in smaller dark matter halos. 

To do this we select only the DLAs from our sample that have low cooling rates, {\elc}. \citet{Wolfe2008} suggests that these DLAs reside in less massive dark matter halos compared to the high cooling rate DLAs. Figure \ref{fig:dvmoddist}a shows the {\delv} distribution of the complete sample (shaded red histogram); the gray solid histogram is the distribution obtained from the simulations by \citet{Pontzen2008}. As can be seen from the inset and the KS test probability, there is a less than $1 \times 10^{-6}$ probability the two distributions are drawn from the same parent population. On the other hand, Figure \ref{fig:dvmoddist}b shows that the DLAs with a low cooling rate (shaded green histogram) have a {\delv} distribution similar to that obtained from the numerical simulation. This result supports the hypothesis that the numerical simulations are capable of reproducing the {\delv} distribution of the DLAs in less massive dark matter halos, but have difficulty producing enough DLAs that reside in more massive dark matter halos. This conclusion was also recently drawn in \citet{Font-Ribera2012} using the cross-correlation between DLAs and the Lyman $\alpha$ forest.

One way numerical simulations try to mitigate the discrepancy in the observed and simulated velocity width distribution is by adding in galactic scale outflows \citep{Tescari2009, Hong2010, Cen2012}. These outflows are capable of decreasing the cross section of less massive dark matter halos, somewhat decoupling the {\delv} statistic from the mass, and increasing the cross section of the more massive dark matter halos. All of these effects increase the relative number of DLAs in more massive dark matter halos. With the addition of galactic scale outflows, these models have been relatively successful in reproducing the {\delv} distribution of DLAs \citep{Tescari2009, Hong2010, Cen2012}, which is supported by the sample in this paper. However, \citet{Font-Ribera2012} suggest that this process still might not produce enough DLAs hosted within massive dark matter halos, indicating that there is still some discrepancy between the observations and the simulations. 

\section{Summary}
\label{sec:summary}
We have studied the spectra of 100 DLAs taken with a single high resolution instrument, the High Resolution Echelle Spectrometer on the Keck telescope, to determine the fundamental relations that exist between the measured parameters (Table \ref{tab:tst}). We find the following new results.
\begin{enumerate}
\item
The metallicity-{\hi} dependency shown in Figure \ref{fig:mhvsnhi} shows a lack of  high {\hi} column density, low metallicity systems besides the known lack of  high {\hi} column density, high metallicity systems \citep{Boisse1998}. Possible explanations for the lack of high {\hi} column density, low metallicity systems are that (1) low metallcity systems arise from cold flows which are not dense enough to form high {\hi} column density systems \citep{Fumagalli2011}, or (2) that higher {\hi} column density systems arise from sightlines probing the inner part of galaxies, which have higher metallicity because of an existing metallicity gradient \citep{Chen2005}. However, neither of these explanations can explain the reflection symmetry that exists about the line [M/H] $=-1.43$. One possible explanation that takes this reflection symmetry as a premise, is that higher {\hi} column density systems probe a larger number of components. The larger number of components results in a smaller probability of encountering a sightline with a very low or very high metallicity. Hence, this will decrease the variance around the mean metallicity value of $-1.43$ for higher column density systems.
\item
The {\delv} parameter is not evolving between redshift 2 through 5 (Figure \ref{fig:dvandwvsz}a). This indicates that at all redshifts, DLAs are hosted by a wide variety of dark matter halo masses, and it allows for a combined description of the {\delv}-metallicity and redshift-metallicity correlations.
\item
The main result of the paper is that we can describe the relations that exist between the redshift, metallicity and mass of a DLA by a single fundamental plane equation: $\rm{[M/H]}=(-1.9{\pm}0.5)+(0.74{\pm}0.21){\cdot}{\log}{\Delta}v_{90}-(0.32{\pm}0.06){\cdot}z$. This plane equation has as advantage that it reduces the scatter around either of the correlations; providing a more stringent constraint for numerical simulations. Secondly, it confirms the existence of a mass-metallicity relationship at each redshift between redshift 2 through 5, where the zero point evolves with redshift. This evolution in the zero point with redshift is consistent with the evolution of the zero point of the mass-metallicity relationship seen in star forming galaxies.
\item
Finally, we compare the sample data to numerical models, and find that numerical models are unable to reproduce the {\delv} distribution of the complete sample (Figure \ref{fig:dvmoddist}a). However, these models are able to reproduce the {\delv} distribution of the low-cool subset of the sample (Figure \ref{fig:dvmoddist}b). These low-cool DLAs are believed to reside in smaller dark matter halos \citep{Wolfe2008}. This result therefore supports the hypothesis that numerical models fail to produce enough DLAs in massive dark matter halos \citep{Pontzen2008, Tescari2009}.
\end{enumerate}

\acknowledgements
Support for this work was provided by NSF grant 11-09447. The data presented herein were obtained at the W.M. Keck Observatory, which is operated as a scientific partnership among the California Institute of Technology, the University of California and the National Aeronautics and Space Administration. The Observatory was made possible by the generous financial support of the W.M. Keck Foundation. The authors wish to recognize and acknowledge the very significant cultural role and reverence that the summit of Mauna Kea has always had within the indigenous Hawaiian community.  We are most fortunate to have the opportunity to conduct observations from this mountain.

\bibliography{Kin_bib}

\end{document}

%% file: table1.tex
\begin{deluxetable*}{lllclcllccccccll}
\tabletypesize{\scriptsize}
\tablecaption{{Journal of Observations}
\label{tab:obs}}
\tablewidth{0pt}
\tablehead{
\colhead{QSO} &
\colhead{RA} &
\colhead{DEC} &
\colhead{$z_{\rm{em}}$\,\tablenotemark{a}} &
\colhead{Date Observed} &
\colhead{Exposure Time} &
\colhead{{\delvres}\,\tablenotemark{b}} &
\colhead{S/N\,\tablenotemark{c}} \\
\colhead{} &
\colhead{(J2000.0)} &
\colhead{(J2000.0)} &
\colhead{} &
\colhead{(UT)} &
\colhead{(s)} &
\colhead{(km s$^{-1}$)} &
\colhead{}
}
\startdata
Q2359$-$02 & 00 01 50.0 & $-$01 59 40.3 & 2.800 & 1997 Sep 29 & 14400 & 8 & 17 &  \\
 & & & & 1997 Sep 30 & 10800 & 8 &  &  \\
 & & & & 1997 Oct 01 & 12516 & 8 &  &  \\
Q0000$-$2619 & 00 03 22.9 & $-$26 03 16.8 & 4.110 & 1994 Sep 30 & 10800 & 8 & 15 &  \\
 & & & & 1994 Oct 01 & 3600 & 8 &  &  \\
BR0019$-$15 & 00 22 08.0 & $-$15 05 38.8 & 4.530 & 1996 Sep 20 & 13500 & 8 & 18 &  \\
 & & & & 1996 Sep 21 & 12600 & 8 &  &  \\
 & & & & 1996 Sep 22 & 14400 & 8 &  &  \\
J0040$-$0915 & 00 40 54.7 & $-$09 19 26.9 & 4.976 & 2011 Jan 16 & 7200 & 6 & 7 &  \\
 & & & & 2011 Jan 24 & 3600 & 6 &  &  
\enddata
\tablecomments{Units of right ascension are in hours, minutes, and seconds, and units of declination are in degrees, arcminutes, and arcseconds.}
\tablenotetext{a}{The emission redshift of the quasar.}
\tablenotetext{b}{{\delvres} is defined as the FWHM resolution of the spectrum.}
\tablenotetext{c}{The average signal to noise (S/N) ratio per 1.4 {\kms} pixel.}
\tablecomments{(This table is available in its entirety in a machine-readable form in the online journal. A portion is shown here for guidance regarding its form and content.)}
\end{deluxetable*}

%% file: table2.tex
\begin{deluxetable*}{llccccccll}
\tabletypesize{\scriptsize}
\tablecaption{{HIRES DLA sample}
\label{tab:dat}}
\tablewidth{0pt}
\tablehead{
\colhead{QSO} &
\colhead{$z_{\rm{abs}}$} &
\colhead{$\log N_{\rm HI}$} &
\colhead{Metallicity} &
\colhead{${\log}$ {\elc}} &
\colhead{{\w1526}} &
\colhead{{\delv}\,\tablenotemark{a}} &
\colhead{Transition\,\tablenotemark{b}} &
\colhead{Selection} &
\colhead{References} \\
\colhead{} &
\colhead{} &
\colhead{(cm$^{-2}$)} &
\colhead{} &
\colhead{(ergs s$^{-1}$ H$^{-1}$)} &
\colhead{(\AA)} &
\colhead{({\kms})} &
\colhead{} &
\colhead{Criterion} &
\colhead{} \\
}
\startdata
Q1104$-$18 & 1.6613 & 20.80 $\pm$ 0.10 & $-0.99$ $\pm$ 0.10 & $-26.87$$\pm$0.11 & --- & 50 & Si{\tiny{ II}}$\lambda$1808 & H{\tiny{ I}} - selected & 6 \\
Q1331$+$17 & 1.7763 & 21.14 $\pm$ 0.08 & $-1.34$ $\pm$ 0.08 & $-26.60$$\pm$0.09 & 0.499$\pm$0.001 & 72 & Si{\tiny{ II}}$\lambda$1808 & H{\tiny{ I}} - selected & 7, 13, 18, 24 \\
Q0841$+$12 & 1.8640 & 21.00 $\pm$ 0.10 & $-1.46$ $\pm$ 0.10 & --- & --- & 30 & Si{\tiny{ II}}$\lambda$1808 & Serendipitous & 20 \\
Q2230$+$02 & 1.8643 & 20.85 $\pm$ 0.08 & $-0.71$ $\pm$ 0.09 & --- & 1.500$\pm$0.018 & 172 & Si{\tiny{ II}}$\lambda$1808 & H{\tiny{ I}} - selected & 7, 13 \\
Q1210$+$17 & 1.8917 & 20.60 $\pm$ 0.10 & $-0.79$ $\pm$ 0.10 & --- & 0.419$\pm$0.003 & 38 & Si{\tiny{ II}}$\lambda$1808 & H{\tiny{ I}} - selected & 13, 19
\enddata
\tablenotetext{a}{The uncertainty on the {\delv} measurement is taken to be 10 {\kms} for each measurement.}
\tablenotetext{b}{Transition used to measure the {\delv} parameter.}
\tablerefs{(1) \citet{Wolfe1994};(2) \citet{lu1996};(3) \citet{Prochaska1996};(4) \citet{Prochaska1997};
(5) \citet{Lu1998}; (6) \citet{Lopez1999}; (7) \citet{Prochaska1999}; (8) \citet{Molaro2000}; (9) \citet{Petitjean2000};
(10) \citet{Prochaska2000}; (11) \citet{Ellison2001a}; (12) \citet{Molaro2001}; (13) \citet{Prochaska2001}; (14) \citet{Ledoux2002};
(15) \citet{Levshakov2002}; (16) \citet{Prochaska2002a}; (17) \citet{Prochaska2003a}; (18) \citet{Dessauges2004}; (19) \citet{Dessauges2006};
(20) \citet{Ledoux2006}; (21) \citet{Dessauges2007}; (22) \citet{Prochaska2007}; (23) \citet{Wolfe2008}; (24) \citet{Jorgenson2010};
(25) \citet{Vladilo2011}; (26) \citet{Rafelski2012}
}
\tablecomments{(This table is available in its entirety in a machine-readable form in the online journal. A portion is shown here for guidance regarding its form and content.)}
\end{deluxetable*}

%% file: table3.tex
\begin{deluxetable}{cccccc}[!b]
\tabletypesize{\scriptsize}
\tablecaption{{Fitting Functions to the Distributions}
\label{tab:fit}}
\tablewidth{0pt}
\tablehead{
\colhead{x} &
\colhead{$\mu_1$} &
\colhead{$\sigma_1^2$} &
\colhead{$\mu_2$} &
\colhead{$\sigma_2^2$} &
\colhead{$r$} \\
}
\startdata
$\log$ {\delv}&$1.83$&$0.39$& --- & --- & --- \\
$\log$ {\w1526}&$-0.44$&$0.42$& --- & --- & --- \\
$[M/H]$&$-1.46$&$0.55$& --- & --- & --- \\
$\log$ {\elc}&$-27.4$&$0.10$&$-26.7$&$0.28$&$0.55$\\
\cutinhead{Function Used}
\vspace{0.5cm}
(a)&\multicolumn{5}{c}{$f(x;\mu_1,\sigma_1^2) =\frac{1}{\sqrt{2\pi\sigma_1^2}}e^{-\frac{(x-\mu_1)^2}{2\sigma_1^2}}$}\\
(b)&\multicolumn{5}{c}{$f(x; \mu_i,\sigma_i^2)=(\frac{r}{1+r})f(x;\mu_1,\sigma_1^2)+(\frac{1}{1+r})f(x;\mu_2,\sigma_2^2)$}
\enddata
\tablecomments{(a) is a normalized Gaussian distribution function with mean, $\mu_1$ and variance $\sigma_1^2$. (b) is the sum of two Gaussian functions. The factors in front of the Gaussian terms are required such that the total integrated area under the function is equal to unity. The r-parameter is the ratio of the relative sizes of the two Gaussian distributions.}
\end{deluxetable}

%% file: table4.tex
\begin{deluxetable*}{lclrrccccc}
\tabletypesize{\scriptsize}
\tablecaption{{Table of Potential Correlations}
\label{tab:tst}}
\tablewidth{0pt}
\tablecolumns{10}
\tablehead{
\multicolumn{3}{c}{Dependency} &
\multicolumn{2}{c}{Linear regression line\,\tablenotemark{a}} &
\colhead{Kendall\,\tablenotemark{b}} &
\colhead{KS-test\,\tablenotemark{b}} &
\colhead{F-test\,\tablenotemark{b}} &
\colhead{U-test\,\tablenotemark{b}} &
\colhead{T-test\,\tablenotemark{b}} \\
\cline{4-5}\\
\colhead{{($y$)}} &
\colhead{} &
\colhead{($x$)} &
\colhead{($a$)} &
\colhead{($b$)} &
\colhead{} &
\colhead{} &
\colhead{} &
\colhead{} &
\colhead{}\\
}
\startdata
$z$&vs&log {\nh}&$-$0.24 $\pm$ 0.26&8. $\pm$ 5.&$0.211$&$0.403$&$0.677$&$0.160$&$0.392$\\
$[M/H]$&vs&log {\nh}&0.04 $\pm$ 0.14&$-$2.3 $\pm$ 2.8&$0.688$&$0.403$&$0.013$&$0.278$&$0.845$\\
log {\delv}&vs&log {\nh}&0.07 $\pm$ 0.10&0.4 $\pm$ 2.1&$0.384$&$0.403$&$0.527$&$0.199$&$0.404$\\
log {\w1526}&vs&log {\nh}&0.21 $\pm$ 0.11&$-$4.8 $\pm$ 2.4&$0.040$&$0.035$&$0.414$&$0.083$&$0.260$\\
log {\elc}&vs&log {\nh}&$-$0.58 $\pm$ 0.21&$-$15. $\pm$ 4.&$0.011$&$0.019$&$0.911$&$0.002$&$0.002$\\
$[M/H]$&vs&$z$&$-$0.30 $\pm$ 0.06&$-$0.53 $\pm$ 0.19&$1.22\rm{E}-5$&$1.65\rm{E}-4$&$0.816$&$3.06\rm{E}-5$&$1.84\rm{E}-5$\\
log {\delv}&vs&$z$&$-$0.00 $\pm$ 0.04&1.84 $\pm$ 0.13&$0.862$&$0.811$&$0.487$&$0.232$&$0.648$\\
log {\w1526}&vs&$z$&$-$0.10 $\pm$ 0.05&$-$0.08 $\pm$ 0.15&$0.042$&$0.563$&$0.190$&$0.069$&$0.125$\\
log {\elc}&vs&$z$&$-$0.05 $\pm$ 0.09&$-$26.83 $\pm$ 0.29&$0.849$&$0.991$&$0.761$&$0.409$&$0.771$\\
log {\delv}&vs&$[M/H]$&0.40 $\pm$ 0.04&2.40 $\pm$ 0.06&$< 1\rm{E}-6$&$< 1\rm{E}-6$&$0.457$&$< 1\rm{E}-6$&$< 1\rm{E}-6$\\
log {\w1526}&vs&$[M/H]$&0.58 $\pm$ 0.06&0.39 $\pm$ 0.08&$< 1\rm{E}-6$&$< 1\rm{E}-6$&$0.464$&$< 1\rm{E}-6$&$< 1\rm{E}-6$\\
log {\elc}&vs&$[M/H]$&0.26 $\pm$ 0.15&$-$26.63 $\pm$ 0.21&$0.027$&$0.065$&$0.411$&$0.028$&$0.082$\\
log {\w1526}&vs&log {\delv}&0.83 $\pm$ 0.07&$-$1.91 $\pm$ 0.13&$< 1\rm{E}-6$&$< 1\rm{E}-6$&$0.603$&$< 1\rm{E}-6$&$< 1\rm{E}-6$\\
log {\elc}&vs&log {\delv}&0.46 $\pm$ 0.25&$-$27.8 $\pm$ 0.5&$0.006$&$0.019$&$0.182$&$0.007$&$0.026$\\
log {\elc}&vs&log {\w1526}&0.59 $\pm$ 0.29&$-$26.76 $\pm$ 0.11&$0.011$&$0.009$&$0.157$&---\,\tablenotemark{c}&$0.046$
\enddata
\tablenotetext{a}{These are the best fit parameters for the linear regression line of the form: $y=ax+b$}
\tablenotetext{b}{Values are probabilities that the null hypothesis of each test can be rejected}
\tablenotetext{c}{Not enough data points are available to get an accurate measurement}
\end{deluxetable*}

%% file: FPoD.bbl
\begin{thebibliography}{94}
\expandafter\ifx\csname natexlab\endcsname\relax\def\natexlab#1{#1}\fi

\bibitem[{Abazajian {et~al.}(2009)Abazajian, Adelman-McCarthy, Ag\"{u}eros,
  Allam, Prieto, An, Anderson, Anderson, Annis, \& Bahcall}]{Abazajian2009}
Abazajian, K.~N., Adelman-McCarthy, J.~K., Ag\"{u}eros, M.~A., {et~al.} 2009,
  ApJS, 182, 543

\bibitem[{Akerman {et~al.}(2005)Akerman, Ellison, Pettini, \&
  Steidel}]{Akerman2005}
Akerman, C., Ellison, S.~L., Pettini, M., \& Steidel, C.~C. 2005, A{\&}A, 440,
  499

\bibitem[{Asplund {et~al.}(2009)Asplund, Grevesse, Sauval, \&
  Scott}]{Asplund2009}
Asplund, M., Grevesse, N., Sauval, A.~J., \& Scott, P. 2009, ARA{\&}A, 47, 481

\bibitem[{Bernardi {et~al.}(2003)Bernardi, Sheth, Annis, Burles, Eisenstein,
  Finkbeiner, Hogg, \& Lupton}]{Bernardi2003}
Bernardi, M., Sheth, R.~K., Annis, J., {et~al.} 2003, AJ, 125, 1866

\bibitem[{Boiss\'{e} {et~al.}(1998)Boiss\'{e}, Le~Brun, Bergeron, \&
  Deharveng}]{Boisse1998}
Boiss\'{e}, P., Le~Brun, V., Bergeron, J., \& Deharveng, J.-M. 1998, A{\&}A,
  333, 841

\bibitem[{Brinchmann \& Ellis(2000)}]{Brinchmann2000}
Brinchmann, J., \& Ellis, R. 2000, ApJ, 536, L77

\bibitem[{Cen(2012)}]{Cen2012}
Cen, R. 2012, ApJ, 748, 121

\bibitem[{Chen {et~al.}(2005)Chen, Kennicutt~Jr, \& Rauch}]{Chen2005}
Chen, H.-W., Kennicutt~Jr, R.~C., \& Rauch, M. 2005, ApJ, 620, 703

\bibitem[{Cooke {et~al.}(2011)Cooke, Pettini, Steidel, \& Rudie}]{Cooke2011}
Cooke, R., Pettini, M., Steidel, C.~C., \& Rudie, G.~C. 2011, MNRAS, 417, 1534

\bibitem[{Dessauges-Zavadsky {et~al.}(2004)Dessauges-Zavadsky, Calura,
  Prochaska, D'Odorico, \& Matteucci}]{Dessauges2004}
Dessauges-Zavadsky, M., Calura, F., Prochaska, J.~X., D'Odorico, S., \&
  Matteucci, F. 2004, A{\&}A, 416, 79

\bibitem[{Dessauges-Zavadsky {et~al.}(2006)Dessauges-Zavadsky, Calura,
  Prochaska, D'Odorico, \& Matteucci}]{Dessauges2006}
---. 2006, A{\&}A, 445, 93

\bibitem[{Dessauges-Zavadsky {et~al.}(2007)Dessauges-Zavadsky, Calura,
  Prochaska, D'Odorico, \& Matteucci}]{Dessauges2007}
---. 2007, A{\&}A, 470, 431

\bibitem[{Djorgovski \& Davis(1987)}]{Djorgovski1987}
Djorgovski, S., \& Davis, M. 1987, ApJ, 313, 59

\bibitem[{Ellison {et~al.}(2005)Ellison, Hall, \& Lira}]{Ellison2005}
Ellison, S.~L., Hall, P.~B., \& Lira, P. 2005, AJ, 130, 1345

\bibitem[{Ellison {et~al.}(2001a)Ellison, Pettini, Steidel, \&
  Shapley}]{Ellison2001a}
Ellison, S.~L., Pettini, M., Steidel, C.~C., \& Shapley, A.~E. 2001a, ApJ, 549,
  770

\bibitem[{Ellison {et~al.}(2010)Ellison, Prochaska, Hennawi, Lopez, Usher,
  Wolfe, Russell, \& Benn}]{Ellison2010}
Ellison, S.~L., Prochaska, J.~X., Hennawi, J., {et~al.} 2010, MNRAS, 406, 1453

\bibitem[{Ellison {et~al.}(2001b)Ellison, Yan, Hook, Pettini, Wall, \&
  Shaver}]{Ellison2001b}
Ellison, S.~L., Yan, L., Hook, I., {et~al.} 2001b, A{\&}A, 379, 393

\bibitem[{Erb {et~al.}(2006)Erb, Shapley, Pettini, Steidel, Reddy, \&
  Adelberger}]{Erb2006}
Erb, D.~K., Shapley, A.~E., Pettini, M., {et~al.} 2006, ApJ, 644, 813

\bibitem[{Font-Ribera {et~al.}(2012)Font-Ribera, Miralda-Escud\'{e}, Arnau,
  Carithers, Lee, Noterdaeme, P\^{a}ris, \& Petitjean}]{Font-Ribera2012}
Font-Ribera, A., Miralda-Escud\'{e}, J., Arnau, E., {et~al.} 2012, JCAP, 11,
  059, arXiv:1209.4596v1

\bibitem[{Fox {et~al.}(2007)Fox, Ledoux, Petitjean, \& Srianand}]{Fox2007}
Fox, A.~J., Ledoux, C., Petitjean, P., \& Srianand, R. 2007, A{\&}A, 473, 791

\bibitem[{Frank \& P\'{e}roux(2010)}]{Frank2010}
Frank, S., \& P\'{e}roux, C. 2010, MNRAS, 406, 2235

\bibitem[{Fumagalli {et~al.}(2011)Fumagalli, Prochaska, Kasen, Dekel, Ceverino,
  \& Primack}]{Fumagalli2011}
Fumagalli, M., Prochaska, J.~X., Kasen, D., {et~al.} 2011, MNRAS, 418, 1796

\bibitem[{Fynbo {et~al.}(2010)Fynbo, Laursen, Ledoux, M{\o}ller, Durgapal,
  Goldoni, \& Gullberg}]{Fynbo2010}
Fynbo, J., Laursen, P., Ledoux, C., {et~al.} 2010, MNRAS, 408, 2128

\bibitem[{Fynbo {et~al.}(2011)Fynbo, Ledoux, Noterdaeme, Christensen,
  M{\o}ller, Durgapal, Goldoni, \& Kaper}]{Fynbo2011}
Fynbo, J., Ledoux, C., Noterdaeme, P., {et~al.} 2011, MNRAS, 413, 2481

\bibitem[{Graves \& Faber(2010)}]{Graves2010}
Graves, G.~J., \& Faber, S.~M. 2010, ApJ, 717, 803

\bibitem[{Haehnelt {et~al.}(1998)Haehnelt, Steinmetz, \& Rauch}]{Haehnelt1998}
Haehnelt, M.~G., Steinmetz, M., \& Rauch, M. 1998, ApJ, 495, 647

\bibitem[{Herbert-Fort {et~al.}(2006)Herbert-Fort, Prochaska,
  Dessauges-Zavadsky, Ellison, Howk, Wolfe, \& Prochter}]{Herbert-Fort2006}
Herbert-Fort, S., Prochaska, J.~X., Dessauges-Zavadsky, M., {et~al.} 2006,
  PASP, 118, 1077

\bibitem[{Hong {et~al.}(2010)Hong, Katz, Dav\'{e}, Fardal, Du\v{s}an, \&
  Oppenheimer}]{Hong2010}
Hong, S., Katz, N., Dav\'{e}, R., {et~al.} 2010, arXiv:1008.4242v2

\bibitem[{Jorgenson {et~al.}(2010)Jorgenson, Wolfe, \&
  Prochaska}]{Jorgenson2010}
Jorgenson, R.~A., Wolfe, A.~M., \& Prochaska, J.~X. 2010, ApJ, 722, 460

\bibitem[{Jorgenson {et~al.}(2006)Jorgenson, Wolfe, Prochaska, Lu, Howk, Cooke,
  Gawiser, \& Gelino}]{Jorgenson2006}
Jorgenson, R.~A., Wolfe, A.~M., Prochaska, J.~X., {et~al.} 2006, ApJ, 646, 730

\bibitem[{Kaplan {et~al.}(2010)Kaplan, Prochaska, Herbert-Fort, Ellison, \&
  Dessauges-Zavadsky}]{Kaplan2010}
Kaplan, K.~F., Prochaska, J.~X., Herbert-Fort, S., Ellison, S.~L., \&
  Dessauges-Zavadsky, M. 2010, MNRAS, 122, 619

\bibitem[{Kormendy(1982)}]{Kormendy1982}
Kormendy, J. 1982, in Proc. 12th Saas-Fe Advanced Course Morphology and
  Dynamics of Galaxies (Sauverny: Geneva Obs. Pub.), 113--288

\bibitem[{Krumholz {et~al.}(2009)Krumholz, Ellison, Prochaska, \&
  Tumlinson}]{Krumholz2009}
Krumholz, M.~R., Ellison, S.~L., Prochaska, J.~X., \& Tumlinson, J. 2009, ApJ,
  701, L12

\bibitem[{Kulkarni {et~al.}(2005)Kulkarni, Fall, Lauroesch, York, Welty, Khare,
  \& Truran}]{Kulkarni2005}
Kulkarni, V.~P., Fall, S.~M., Lauroesch, J.~T., {et~al.} 2005, ApJ, 618, 68

\bibitem[{Ledoux {et~al.}(1998)Ledoux, Petitjean, Bergeron, Wampler, \&
  Srianand}]{Ledoux1998}
Ledoux, C., Petitjean, P., Bergeron, J., Wampler, E.~J., \& Srianand, R. 1998,
  A{\&}A, 337, 51

\bibitem[{Ledoux {et~al.}(2006)Ledoux, Petitjean, Fynbo, M{\o}ller, \&
  Srianand}]{Ledoux2006}
Ledoux, C., Petitjean, P., Fynbo, J., M{\o}ller, P., \& Srianand, R. 2006,
  A{\&}A, 457, 71

\bibitem[{Ledoux {et~al.}(2002)Ledoux, Srianand, \& Petitjean}]{Ledoux2002}
Ledoux, C., Srianand, R., \& Petitjean, P. 2002, A{\&}A, 392, 781

\bibitem[{Levshakov {et~al.}(2002)Levshakov, Dessauges-Zavadsky, D'Odorico, \&
  Molaro}]{Levshakov2002}
Levshakov, S.~A., Dessauges-Zavadsky, M., D'Odorico, S., \& Molaro, P. 2002,
  ApJ, 565, 696

\bibitem[{Lopez {et~al.}(1999)Lopez, Reimers, Rauch, Sargent, \&
  Smette}]{Lopez1999}
Lopez, S., Reimers, D., Rauch, M., Sargent, W.~L., \& Smette, A. 1999, ApJ,
  513, 598

\bibitem[{Lu {et~al.}(1996)Lu, Sargent, \& Barlow}]{lu1996}
Lu, L., Sargent, W.~L., \& Barlow, T.~A. 1996, ApJ Supp., 107, 475

\bibitem[{Lu {et~al.}(1998)Lu, Sargent, \& Barlow}]{Lu1998}
---. 1998, AJ, 115, 55

\bibitem[{Maiolino {et~al.}(2008)Maiolino, Nagao, Grazian, Cocchia, Marconi,
  Mannucci, Cimatti, Pipino, \& Banero}]{Maiolino2008}
Maiolino, R., Nagao, T., Grazian, A., {et~al.} 2008, A{\&}A, 488, 463

\bibitem[{Markwardt(2009)}]{Markwardt2009}
Markwardt, C.~B. 2009, ADASS, 411, 251

\bibitem[{Merloni {et~al.}(2003)Merloni, Heinz, \& Di~Matteo}]{Merloni2003}
Merloni, A., Heinz, S., \& Di~Matteo, T. 2003, MNRAS, 345, 1057

\bibitem[{Mo {et~al.}(1998)Mo, Mao, \& White}]{Mo1998}
Mo, H.~J., Mao, S., \& White, S.~D. 1998, MNRAS, 295, 319

\bibitem[{Molaro {et~al.}(2000)Molaro, Bonifacio, Centuri\'{o}n, D'Odorico,
  Vladilo, Santin, \& Di~Marcantonio}]{Molaro2000}
Molaro, P., Bonifacio, P., Centuri\'{o}n, M., {et~al.} 2000, ApJ, 541, 54

\bibitem[{Molaro {et~al.}(2001)Molaro, Levshakov, D'Odorico, Bonifacio, \&
  Centuri\'{o}n}]{Molaro2001}
Molaro, P., Levshakov, S.~A., D'Odorico, S., Bonifacio, P., \& Centuri\'{o}n,
  M. 2001, ApJ, 549, 90

\bibitem[{M{\o}ller {et~al.}(2013)M{\o}ller, Fynbo, Ledoux, \&
  Nilsson}]{Moller2013}
M{\o}ller, P., Fynbo, J., Ledoux, C., \& Nilsson, K. 2013, MNRAS, In Press

\bibitem[{Morton(2003)}]{Morton2003}
Morton, D.~C. 2003, ApJS, 149, 205

\bibitem[{Moustakas {et~al.}(2010)Moustakas, Brown, Cool, Dey, Eisenstein,
  Gonzalez, Jannuzi, Jones, Kochanek, Murray, \& Wild}]{Moustakas2012}
Moustakas, J., Brown, M., Cool, R.~J., {et~al.} 2010, arXiv:astro-ph/1112.3300

\bibitem[{Noterdaeme {et~al.}(2009)Noterdaeme, Petitjean, Ledoux, \&
  Srianand}]{Noterdaeme2009}
Noterdaeme, P., Petitjean, P., Ledoux, C., \& Srianand, R. 2009, A{\&}A, 505,
  1087

\bibitem[{Noterdaeme {et~al.}(2012)Noterdaeme, Petitjean, Carithers, P\^{a}ris,
  Font-Ribera, Bailey, Auborg, Bizyaev, Ebelke, \& Finley}]{Noterdaeme2012}
Noterdaeme, P., Petitjean, P., Carithers, W., {et~al.} 2012, A{\&}A, 547, L1

\bibitem[{Penprase {et~al.}(2010)Penprase, Prochaska, Sargent, Toro-Martinez,
  \& Beeler}]{Penprase2010}
Penprase, B.~E., Prochaska, J.~X., Sargent, W.~L., Toro-Martinez, I., \&
  Beeler, D.~J. 2010, ApJ, 721, 1

\bibitem[{Petitjean {et~al.}(2000)Petitjean, Srianand, \&
  Ledoux}]{Petitjean2000}
Petitjean, P., Srianand, R., \& Ledoux, C. 2000, A{\&}A, 364, L26

\bibitem[{Petitjean {et~al.}(2002)Petitjean, Srianand, \&
  Ledoux}]{Petitjean2002}
---. 2002, MNRAS, 332, 383

\bibitem[{Pettini {et~al.}(1994)Pettini, Smith, Hunstead, \&
  King}]{Pettini1994}
Pettini, M., Smith, L.~J., Hunstead, R.~W., \& King, D.~L. 1994, ApJ, 426, 79

\bibitem[{Pettini {et~al.}(2008)Pettini, Zych, Steidel, \&
  Chaffee}]{Pettini2008}
Pettini, M., Zych, B.~J., Steidel, C.~C., \& Chaffee, F.~H. 2008, MNRAS, 385,
  2011

\bibitem[{Pontzen {et~al.}(2008)Pontzen, Governato, Pettini, Booth, Stinson,
  Wadsley, Brooks, Quinn, \& Haehnelt}]{Pontzen2008}
Pontzen, A., Governato, F., Pettini, M., {et~al.} 2008, MNRAS, 390, 1349

\bibitem[{Pottasch {et~al.}(1979)Pottasch, Wesselius, \& van
  Duinen}]{Pottasch1979}
Pottasch, S., Wesselius, P., \& van Duinen, R. 1979, A{\&}A, 74, L15

\bibitem[{Prochaska {et~al.}(2008)Prochaska, Chen, Wolfe, Dessauges-Zavadsky,
  \& Bloom}]{Prochaska2008}
Prochaska, J.~X., Chen, H.-W., Wolfe, A.~M., Dessauges-Zavadsky, M., \& Bloom,
  J.~S. 2008, ApJ, 672, 59

\bibitem[{Prochaska {et~al.}(2003a)Prochaska, Gawiser, Wolfe, Castro, \&
  Djorgovski}]{Prochaska2003a}
Prochaska, J.~X., Gawiser, E., Wolfe, A.~M., Castro, S., \& Djorgovski, S.
  2003a, ApJ, 595, L9

\bibitem[{Prochaska {et~al.}(2003b)Prochaska, Gawiser, Wolfe, Cooke, \&
  Gelino}]{Prochaska2003b}
Prochaska, J.~X., Gawiser, E., Wolfe, A.~M., Cooke, J., \& Gelino, D. 2003b,
  ApJS, 147, 227

\bibitem[{Prochaska {et~al.}(2002)Prochaska, Henry, O'Meara, Tytler, Wolfe,
  Kirkman, Lubin, \& Suzuki}]{Prochaska2002a}
Prochaska, J.~X., Henry, R.~B., O'Meara, J.~M., {et~al.} 2002, PASP, 114, 933

\bibitem[{Prochaska {et~al.}(2005)Prochaska, Herbert-Fort, \&
  Wolfe}]{Prochaska2005}
Prochaska, J.~X., Herbert-Fort, S., \& Wolfe, A.~M. 2005, ApJ, 635, 123

\bibitem[{Prochaska \& Wolfe(1996)}]{Prochaska1996}
Prochaska, J.~X., \& Wolfe, A.~M. 1996, ApJ, 470, 403

\bibitem[{Prochaska \& Wolfe(1997)}]{Prochaska1997}
---. 1997, ApJ, 487, 73

\bibitem[{Prochaska \& Wolfe(1999)}]{Prochaska1999}
---. 1999, ApJ Supp., 121, 369

\bibitem[{Prochaska \& Wolfe(2000)}]{Prochaska2000}
---. 2000, ApJ, 533, L5

\bibitem[{Prochaska \& Wolfe(2009)}]{Prochaska2009}
---. 2009, ApJ, 696, 1543

\bibitem[{Prochaska \& Wolfe(2010)}]{Prochaska2010}
---. 2010, arXiv:astro-ph/1009.3960

\bibitem[{Prochaska {et~al.}(2007)Prochaska, Wolfe, Howk, Gawiser, Burles, \&
  Cooke}]{Prochaska2007}
Prochaska, J.~X., Wolfe, A.~M., Howk, J.~C., {et~al.} 2007, ApJS, 171, 29

\bibitem[{Prochaska {et~al.}(2001)Prochaska, Wolfe, Tytler, Burles, Cooke,
  Gawiser, Kirkman, O'Meara, \& Storrie-Lombardi}]{Prochaska2001}
Prochaska, J.~X., Wolfe, A.~M., Tytler, D., {et~al.} 2001, ApJ Supp., 137, 21

\bibitem[{Rafelski {et~al.}(2012)Rafelski, Wolfe, Prochaska, Neeleman, \&
  Mendez}]{Rafelski2012}
Rafelski, M., Wolfe, A.~M., Prochaska, J.~X., Neeleman, M., \& Mendez, A. 2012,
  ApJ, 755, 89

\bibitem[{Razoumov {et~al.}(2006)Razoumov, Norman, Prochaska, \&
  Wolfe}]{Razoumov2006}
Razoumov, A.~O., Norman, M.~L., Prochaska, J.~X., \& Wolfe, A.~M. 2006, ApJ,
  645, 55

\bibitem[{Rea {et~al.}(2012)Rea, Pons, Torres, \& Turolla}]{Rea2012}
Rea, N., Pons, J.~A., Torres, D.~F., \& Turolla, R. 2012, ApJ, 748, L12

\bibitem[{Savage \& Sembach(1991)}]{Savage1991}
Savage, B.~D., \& Sembach, K.~R. 1991, ApJ, 379, 245

\bibitem[{Savaglio {et~al.}(2005)Savaglio, Glazebrook, Le~Borgne, Juneau,
  Abraham, Chen, Crampton, McCarthy, Carlberg, Marzke, Roth, J{\o}rgensen, \&
  Murowinski}]{Savaglio2005}
Savaglio, S., Glazebrook, D., Le~Borgne, D., {et~al.} 2005, ApJ, 635, 260

\bibitem[{Schaye(2001)}]{Schaye2001b}
Schaye, J. 2001, ApJ, 562, L95

\bibitem[{Sheinis {et~al.}(2002)Sheinis, Bolte, Epps, Kibrick, Miller, Radovan,
  Bigelow, \& Sutin}]{Sheinis2002}
Sheinis, A., Bolte, M., Epps, H., {et~al.} 2002, PASP, 114, 851

\bibitem[{Sheth \& Bernardi(2012)}]{Sheth2012}
Sheth, R.~K., \& Bernardi, M. 2012, MNRAS, 422, 1825

\bibitem[{Suzuki {et~al.}(2003)Suzuki, Tytler, Kirkman, O'Meara, \&
  Lubin}]{Suzuki2003}
Suzuki, N., Tytler, D., Kirkman, D., O'Meara, J.~M., \& Lubin, D. 2003, PASP,
  115, 1050

\bibitem[{Tescari {et~al.}(2009)Tescari, Viel, Tornatore, \&
  Borgani}]{Tescari2009}
Tescari, E., Viel, M., Tornatore, L., \& Borgani, S. 2009, MNRAS, 397, 411

\bibitem[{Tremonti {et~al.}(2004)Tremonti, Heckman, Kaufmann, Brinchmann,
  Charlot, White, Seibert, Peng, Schlegel, Uomoto, Fukugita, \&
  Brinkmann}]{Tremonti2004}
Tremonti, C.~A., Heckman, T.~M., Kaufmann, G., {et~al.} 2004, ApJ, 613, 898

\bibitem[{Vladilo {et~al.}(2011)Vladilo, Abate, Yin, Cescutti, \&
  Matteucci}]{Vladilo2011}
Vladilo, G., Abate, C., Yin, J., Cescutti, G., \& Matteucci, F. 2011, A{\&}A,
  530, 33

\bibitem[{Vogt {et~al.}(1994)Vogt, Allen, Bigelow, Bresee, Brown, Cantrall,
  Conrad, Couture, Delaney, \& Epps}]{Vogt1994}
Vogt, S., Allen, S., Bigelow, B., {et~al.} 1994, Proc. SPIE, 2198, 362

\bibitem[{Williams {et~al.}(2010)Williams, Bureau, \&
  Cappellari}]{Williams2010}
Williams, M.~J., Bureau, M., \& Cappellari, M. 2010, MNRAS, 409, 1330

\bibitem[{Wolfe {et~al.}(1994)Wolfe, Fan, Tytler, Vogt, Keane, \&
  Lanzetta}]{Wolfe1994}
Wolfe, A.~M., Fan, X.-M., Tytler, D., {et~al.} 1994, ApJ, 435, L101

\bibitem[{Wolfe {et~al.}(2005)Wolfe, Gawiser, \& Prochaska}]{Wolfe2005}
Wolfe, A.~M., Gawiser, E., \& Prochaska, J.~X. 2005, ARA{\&}A, 43, 861

\bibitem[{Wolfe {et~al.}(1995)Wolfe, Lanzetta, Foltz, \& Chaffee}]{Wolfe1995}
Wolfe, A.~M., Lanzetta, K.~A., Foltz, C.~B., \& Chaffee, F.~H. 1995, ApJ, 454,
  698

\bibitem[{Wolfe \& Prochaska(1998)}]{Wolfe1998}
Wolfe, A.~M., \& Prochaska, J.~X. 1998, ApJ, 494, L15

\bibitem[{Wolfe \& Prochaska(2000)}]{Wolfe2000}
---. 2000, ApJ, 545, 591

\bibitem[{Wolfe {et~al.}(2003)Wolfe, Prochaska, \& Gawiser}]{Wolfe2003}
Wolfe, A.~M., Prochaska, J.~X., \& Gawiser, E. 2003, ApJ, 593, 215

\bibitem[{Wolfe {et~al.}(2008)Wolfe, Prochaska, Jorgenson, \&
  Rafelski}]{Wolfe2008}
Wolfe, A.~M., Prochaska, J.~X., Jorgenson, R.~A., \& Rafelski, M. 2008, ApJ,
  681, 881

\bibitem[{Wright {et~al.}(1991)Wright, Mather, Bennett, Cheng, Shafer, Fixsen,
  Eplee, \& Isaacman}]{Wright1991}
Wright, E., Mather, J., Bennett, C., {et~al.} 1991, ApJ, 381, 200

\end{thebibliography}
